\documentclass[twocolumn, usenames]{aastex631}
\usepackage{CJK}
\usepackage{amsmath}
\usepackage{inputenc}
\CJKspace
\usepackage{longtable}

\newcommand{\msun}{M\ensuremath{_{\odot}}}

\newcommand{\rsun}[1]{$\mathrm{R_{\odot}}$}
\newcommand{\lsun}[1]{$\mathrm{L_{\odot}}$}
\newcommand{\Halpha}[1]{$\mathrm{H\alpha}$}
\newcommand{\Hbeta}[1]{$\mathrm{H\beta}$}

\newcommand{\NC}{\ion{N}{3} ($\lambda\lambda4634.0, 4640.6$)/\ion{C}{3} ($\lambda\lambda4647.5, 4650.0$)}
\newcommand{\HeII}{\ion{He}{2} ($\lambda4685.5$)}
\newcommand{\mlunit}{$\mathrm{M_{\odot}\,yr^{-1}}$}

\defcitealias{astropy_collaboration_astropy_2013}{Astropy Collaboration 2013}
\defcitealias{astropy_collaboration_astropy_2018}{2018}
\defcitealias{astropy_collaboration_astropy_2022}{2022}

\usepackage{color}
\newcommand\N[1]{{\color{orange} \bf #1}} 
\usepackage{soul}

\newcommand{\LCO}{\affiliation{Las Cumbres Observatory, 6740 Cortona Drive, Suite 102, Goleta, CA 93117-5575, USA}}
\newcommand{\UCSB}{\affiliation{Department of Physics, University of California, Santa Barbara, CA 93106-9530, USA}}

\newcommand{\UCD}{\affiliation{Department of Physics and Astronomy, University of California, Davis, 1 Shields Avenue, Davis, CA 95616-5270, USA}}

\newcommand{\OAPD}{\affiliation{INAF-Osservatorio Astronomico di Padova, Vicolo dell'Osservatorio 5, I-35122 Padova, Italy}}
\newcommand{\OAB}{\affiliation{INAF-Osservatorio Astronomico di Brera, Via E. Bianchi 46, I-23807, Merate (LC), Italy}}

\newcommand{\STScI}{\affiliation{Space Telescope Science Institute, 3700 San Martin Drive, Baltimore, MD 21218-2410, USA}}
\newcommand{\UT}{\affiliation{University of Texas at Austin, 1 University Station C1400, Austin, TX 78712-0259, USA}}

\newcommand{\CfA}{\affiliation{Center for Astrophysics \textbar{} Harvard \& Smithsonian, 60 Garden Street, Cambridge, MA 02138-1516, USA}}
\newcommand{\UA}{\affiliation{Steward Observatory, University of Arizona, 933 North Cherry Avenue, Tucson, AZ 85721-0065, USA}}

\newcommand{\TAU}{\affiliation{School of Physics and Astronomy, Tel Aviv University, Tel Aviv 69978, Israel}}

\newcommand{\Potsdam}{\affiliation{Leibniz-Institut f\"ur Astrophysik Potsdam (AIP), An der Sternwarte 16, D-14482 Potsdam, Germany}}

\newcommand{\UNC}{\affiliation{Department of Physics and Astronomy, University of North Carolina, 120 East Cameron Avenue, Chapel Hill, NC 27599, USA}}

\newcommand{\MMT}{\affiliation{MMT and Steward Observatories, University of Arizona, 933 North Cherry Avenue, Tucson, AZ 85721-0065, USA}}

\newcommand{\JHU}{\affiliation{Department of Physics and Astronomy, The Johns Hopkins University, 3400 North Charles Street, Baltimore, MD 21218, USA}}

\newcommand{\GeminiNorth}{\affiliation{Gemini Observatory, 670 North A`ohoku Place, Hilo, HI 96720-2700, USA}}
\newcommand{\Keck}{\affiliation{W.~M.~Keck Observatory, 65-1120 M\=amalahoa Highway, Kamuela, HI 96743-8431, USA}}
\newcommand{\UW}{\affiliation{Department of Astronomy, University of Washington, 3910 15th Avenue NE, Seattle, WA 98195-0002, USA}}
\newcommand{\Catalyst}{\altaffiliation{LSSTC Catalyst Fellow}}
\newcommand{\USask}{\affiliation{Department of Physics and Engineering Physics, University of Saskatchewan, 116 Science Place, Saskatoon, SK S7N 5E2, Canada}}

\newcommand{\Rutgers}{\affiliation{Department of Physics and Astronomy, Rutgers, the State University of New Jersey,\\136 Frelinghuysen Road, Piscataway, NJ 08854-8019, USA}}

\newcommand{\ICE}{\affiliation{Institute of Space Sciences (ICE, CSIC), Campus UAB, Carrer
de Can Magrans, s/n, E-08193 Barcelona, Spain}}
\newcommand{\IEEC}{\affiliation{Institut d'Estudis Espacials de Catalunya, Gran Capit\`a, 2-4, Edifici Nexus, Desp.\ 201, E-08034 Barcelona, Spain}}

\newcommand{\Hobart}{\affiliation{Physics Department, Hobart and William Smith Colleges, 300 Pulteney Street, Geneva, NY 14456, USA}}
\newcommand{\Cornell}{\affiliation{Department of Astronomy, Cornell University, 245 East Avenue, Ithaca, NY 14850, USA}}
\newcommand{\Athens}{\affiliation{IAASARS, National Observatory of Athens, Penteli 15236, Greece}}

\newcommand{\UMNAstro}{\affiliation{School of Physics and Astronomy, University of Minnesota, 116 Church Street S.E., Minneapolis, MN 55455, USA}}

\newcommand{\UTA}{\affiliation{Department of Physics, University of Texas at Arlington, Box 19059, Arlington, TX 76019, USA}}
\newcommand{\Konkoly}{\affiliation{Konkoly Observatory, CSFK, MTA Center of Excellence, Konkoly-Thege M. \'ut 15-17, Budapest, 1121, Hungary}}
\newcommand{\ELTE}{\affiliation{ELTE E\"otv\"os Lor\'and University, Institute of Physics and Astronomy, P\'azm\'any P\'eter s\'et\'any 1/A, Budapest, 1117 Hungary}}
\newcommand{\Szeged}{\affiliation{Department of Experimental Physics, University of Szeged, D\'om t\'er 9, Szeged, 6720, Hungary}}
\newcommand{\IAIFI}{\affiliation{The NSF AI Institute for Artificial Intelligence and Fundamental Interactions, USA}}
\newcommand{\UPadua}{\affiliation{Physics and Astronomy Department Galileo Galilei, University of Padova, Vicolo dell'Osservatorio 3, I-35122, Padova, Italy}}
\newcommand{\UWarwick}{\affiliation{Department of Physics, Gibbet Hill Road, University of Warwick, Coventry CV4 7AL, United Kingdom}}
\newcommand{\ING}{\affiliation{Isaac Newton Group of Telescopes, Apt. de Correos 368, E-38700 Santa Cruz de la Palma, Spain}}
\newcommand{\UNott}{\affiliation{School of Physics and Astronomy, University of Nottingham, University Park, Nottingham, NG7 2RD}}
\newcommand{\USurrey}{\affiliation{Mullard Space Science Laboratory, University College London, Holmbury St Mary, Dorking, Surrey RH5 6NT, United Kingdom}}
\newcommand{\USheffiled}{\affiliation{Department of Physics and Astronomy, University of Sheffield, Sheffield S3 7RH, UK}}

\begin{document}

\title{Early Spectroscopy and Dense Circumstellar Medium Interaction in SN~2023ixf}

\correspondingauthor{K. Azalee Bostroem}
\email{bostroem@arizona.edu}
\author[0000-0002-4924-444X]{K.\ Azalee Bostroem}
\Catalyst\UA
\author[0000-0002-0744-0047]{Jeniveve Pearson}
\UA
\author[0000-0002-4022-1874]{Manisha Shrestha}
\UA
\author[0000-0003-4102-380X]{David J.\ Sand}
\UA
\author[0000-0001-8818-0795]{Stefano Valenti}
\UCD
\author[0000-0001-8738-6011]{Saurabh W.\ Jha}
\Rutgers
\author[0000-0003-0123-0062]{Jennifer E.\ Andrews}
\GeminiNorth
\author[0000-0001-5510-2424]{Nathan Smith}
\UA
\author[0000-0003-0794-5982]{Giacomo Terreran}
\LCO
\author{Elizabeth Green}
\UA
\author[0000-0002-7937-6371]{Yize Dong \begin{CJK*}{UTF8}{gbsn}(董一泽)\end{CJK*}}
\UCD
\author[0000-0001-9589-3793]{Michael Lundquist}
\Keck
\author[0000-0002-6703-805X]{Joshua Haislip}
\UNC
\author[0000-0003-2744-4755]{Emily T. Hoang}
\UCD
\author[0000-0002-0832-2974]{Griffin Hosseinzadeh}
\UA
\author[0000-0003-0549-3281]{Daryl Janzen}
\USask
\author[0000-0001-5754-4007]{Jacob E.\ Jencson}
\JHU
\author[0000-0003-3642-5484]{Vladimir Kouprianov}
\UNC
\author[0000-0003-2814-4383]{Emmy Paraskeva}
\UCD
\author[0000-0002-7015-3446]{Nicolas E.\ Meza Retamal}
\UCD
\author[0000-0002-5060-3673]{Daniel E.\ Reichart}
\UNC

\author[0000-0001-7090-4898]{Iair Arcavi}
\TAU
\author[0000-0003-2851-1905]{Alceste Z. Bonanos}
\Athens
\author[0000-0002-8262-2924]{Michael W. Coughlin}
\UMNAstro
\author[0000-0001-5903-8159]{Ross Dobson}
\USurrey\ING
\author[0000-0003-4914-5625]{Joseph Farah}
\LCO\UCSB
\author[0000-0002-1296-6887]{Llu\'is Galbany}
\ICE
\IEEC
\author[0000-0003-2375-2064]{Claudia Guti\'errez}
\IEEC
\ICE
\author{Suzanne Hawley}
\UW
\author[0000-0003-1263-8637]{Leslie Hebb}
\Hobart\Cornell
\author[0000-0002-1125-9187]{Daichi Hiramatsu}
\CfA\IAIFI
\author[0000-0003-4253-656X]{D.\ Andrew Howell}
\LCO\UCSB
\author{Takashi Iijima}
\OAPD
\author[0000-0002-0551-046X]{Ilya Ilyin}
\Potsdam
\author[0000-0000-0000-0000]{Kiran Jhass}
\USheffiled\ING
\author[0000-0001-5807-7893]{Curtis McCully}
\LCO\UCSB
\author[0000-0002-9194-5071]{Sean Moran}
\CfA
\author[0000-0003-2528-3409]{Brett M. Morris}
\STScI
\author[0009-0009-5174-7765]{Alessandra C. Mura}
\UPadua
\author[0000-0003-3939-7167]{Tom\'as E. M\"uller-Bravo}
\ICE
\IEEC
\author[0000-0000-0000-0000]{James Munday}
\UWarwick\ING
\author[0000-0001-9570-0584]{Megan Newsome}
\LCO\UCSB
\author[0009-0004-1807-3053]{Maria Th. Pabst}
\UPadua
\author[0000-0001-5578-8614]{Paolo Ochner}
\OAPD
\UPadua
\author[0000-0003-0209-9246]{Estefania Padilla Gonzalez}
\LCO\UCSB
\author[0000-0002-7259-4624]{Andrea Pastorello}
\OAPD
\author[0000-0002-7472-1279]{Craig Pellegrino}
\LCO\UCSB
\author[0009-0006-4637-4085]{Lara Piscarreta}
\ICE
\author[0000-0002-7352-7845]{Aravind P. Ravi}
\UTA
\author[0000-0003-4254-2724]{Andrea Reguitti}
\OAB
\OAPD
\author[0000-0001-5473-6871]{Laura Salo}
\UMNAstro
\author[0000-0001-8764-7832]{J\'ozsef Vink\'o}
\UT\Konkoly\ELTE\Szeged
\author[0000-0000-0000-0000]{Kellie de Vos}
\UNott\ING
\author[0000-0003-1349-6538]{J. C. Wheeler}
\UT
\author[0000-0002-3452-0560]{G. Grant Williams}
\UA\MMT
\author[0000-0003-2732-4956]{Samuel Wyatt}
\UW

\begin{abstract}

We present the optical spectroscopic evolution of SN~2023ixf seen in sub-night cadence spectra from 1.18 to 15 days after explosion.
We identify high-ionization emission features, signatures of interaction with material surrounding the progenitor star, that fade over the first 7 days, with rapid evolution between spectra observed within the same night.
We compare the emission lines present and their relative strength to those of other supernovae with early interaction, finding a close match to SN~2020pni and SN~2017ahn in the first spectrum and SN~2014G at later epochs.
To physically interpret our observations we compare them to CMFGEN models with confined, dense circumstellar material around a red supergiant progenitor from the literature. 
We find that very few models reproduce the blended \NC{} emission lines observed in the first few spectra and their rapid disappearance thereafter, making this a unique diagnostic. 
From the best models, we find a mass-loss rate of $10^{-3}-10^{-2}$ \mlunit{}, which far exceeds the mass-loss rate for any steady wind, especially for a red supergiant in the initial mass range of the detected progenitor.
These mass-loss rates are, however, similar to rates inferred for other supernovae with early circumstellar interaction.
Using the phase when the narrow emission features disappear, we calculate an outer dense radius of circumstellar material $R_\mathrm{CSM, out}\approx5\times10^{14}~\mathrm{cm}$ and a mean circumstellar material density of $\rho=5.6\times10^{-14}~\mathrm{g\,cm^{-3}}$.
This is consistent with the lower limit on the outer radius of the circumstellar material we calculate from the peak \Halpha{} emission flux, $R_\text{CSM, out}\gtrsim9\times10^{13}~\mathrm{cm}$. 

\end{abstract}


\section{Introduction} \label{sec:intro}
Type II supernovae (SNe; hydrogen-rich, specifically Type IIP/L) are thought to come from red supergiant (RSG) progenitors, with masses of $\sim$8--25~\msun{} \citep{2009smartt, 2015smartt}.
While there is a consensus that massive stars enrich their environments through mass loss, there is no model that quantitatively predicts observed RSG mass-loss rates \citep[][and references therein]{2021kee}. 
Empirical mass-loss rates derived from direct observations span orders of magnitude \citep{2011mauron}.
While the empirical prescription most often used in single-star evolutionary models is that of \citet{1988dejaeger} (which spans 10$^{-7}$--10$^{-3.8}$~\mlunit{} for RSG luminosities of $\log(L/L_\odot) =$ 3.9--5.8) recent analyses have found evidence of significantly higher \citep{2012ekstrom, 2023massey} and lower \citep{2020beasor} mass-loss rates. 
Recently, observations of both the early light curves and spectra of Type II SNe show evidence of dense circumstellar material \citep[CSM;][]{Khazov16, 2018morozova, Bruch22, 2023subrayan}, indicating more extreme mass loss (e.g. eruptive mass loss or a superwind), which are known to occur in more massive stars \citep[e.g. luminous blue variables; LBVs][]{smith11}.
Regardless of the mechanism, all massive stars lose mass, and therefore we expect the photons and ejecta from all of their resultant SN explosions to interact with the circumstellar material (CSM) surrounding the progenitor at some level \citep[see][for a review]{Smith14}.

One of the signatures of CSM interaction in core-collapse (CC) SNe is narrow emission lines corresponding to highly-ionized species in their early spectra. 
Narrow emission lines can first occur when photons from shock breakout ionize surrounding CSM \citep[e.g.][]{Yaron17}.
As the shock passes through the CSM, the kinetic energy of the ejecta is converted to high-energy photons that can also ionize the CSM ahead of the photosphere \citep[e.g.][]{2000leonard, 2015smith, Terreran22}.
The recombination of the ionized gas leads to emission features, with the photons scattering off of the ionized electrons to produce Lorentzian line wings \citep{2001chugai}.
This produces high-ionization features which are a function of the temperature, density, and composition of the CSM (e.g. \ion{O}{6}, \ion{O}{5}, \ion{N}{5}, \ion{N}{4}, \ion{C}{4}, \ion{He}{2}).
As the CSM cools, higher ionization features give way to lower ionization features (e.g. \ion{N}{3}, \ion{O}{3}) and eventually all emission lines fade. 
At the same time, narrow P~Cygni profiles can develop if the CSM is dense enough and sufficiently cool (e.g. \citealt{2000leonard, 2016benetti, Terreran22}).
These profiles can develop into intermediate width features as the CSM begins to be accelerated by the shock.
Eventually, the ejecta sweep up or engulf the CSM and the spectrum begins to develop as a normal CCSN with broad P~Cygni profiles, often with a shallow absorption component.
However, even at this phase the CSM interaction can contribute to the light curve \citep{2022dessart, 2018andrews, 2015smith, 2017smith}.

Analyses of samples of CCSNe with early spectroscopic observations show that a significant fraction of nearby SNe display these features \citep{Khazov16,Bruch21,Bruch22}.  
Detailed modeling of these flash features can constrain the progenitor mass-loss rate just prior to explosion, the surface chemical composition, as well as the extent of the confined CSM \citep[e.g.][]{2017dessart,2019boian,Boian20}.  
There are now dozens of examples of early flash spectroscopy, and several cases where the observations have been modeled or been compared to models in some detail \citep[e.g.][]{Yaron17,Boian20,Tartaglia21,Terreran22,Jacobson22}, but the time evolution of the flash ionization lines has rarely been captured due to their ephemeral nature.

SN~2023ixf was discovered in M101 \citep[D=6.85 Mpc,][]{2022riess} by Koichi Itakagi on 2023-05-19 17:27:15 UTC (all times given in this paper are in UTC; MJD 60083.72) at a magnitude of 14.9 AB mag in a Clear filter \citep{itagaki_discovery_2023}. 
It was classified on 2023-05-19 23:35:34 (MJD 60083.98) as a Type II SN with flash ionization features (H, He, C, and N), using a spectrum taken a few hours after discovery \citep{perley_classification_2023}.
Over the first $\sim$5~days, it rapidly rose to a plateau brightness of $V\approx$ 11.2 mag, or $M_V\approx-18.2$ at the distance to M101 -- a similar brightness to the well-studied Type IIP SN~2004et in NGC~6946 \citep[e.g.][]{Maguire10}.
The early photometric evolution is detailed in a companion paper, by \citet{2023hosseinzadeh}.

In this paper, we present the remarkable early spectroscopic evolution of SN~2023ixf with flash features observed in extraordinary detail.
In \autoref{sec:SpecObs}, we describe our spectroscopic observations, while in \autoref{sec:params}  we present basic properties of SN~2023ixf relevant for our work.  From there, in \autoref{sec:SpecEvolve} we discuss the fast spectroscopic evolution of SN~2023ixf and compare it with existing observational data sets.  In \autoref{sec:models}, we compare our unprecedented flash spectroscopic sequence to existing radiative transfer models to infer the mass-loss rate of the progenitor star and in \autoref{sec:CSM}, we use the spectroscopic evolution to further characterize the CSM of SN~2023ixf and place it in the context of other interacting SNe.  
We summarize and conclude in \autoref{sec:Conclusion}.

\section{Spectroscopic Observations}\label{sec:SpecObs}
Immediately after the discovery announcement, we began a high-cadence, comprehensive campaign to observe the detailed evolution of SN 2023ixf with the Arizona Transient Exploration and Characterization (AZTEC) collaboration, Distance Less Than 40 Mpc (DLT40) collaboration, and the Global Supernova Project.
We observed SN~2023ixf using the moderate-resolution optical spectrograph Hectospec \citep{Fabricant_2005} on the Multiple Mirror Telescope (MMT) on Mt. Hopkins, AZ from 2023-05-20 to 2023-05-26. 
The observed spectra were reduced using an IDL pipeline called HSRED\footnote{\url{http://mingus.as.arizona.edu/~bjw/mmt/hecto_reduction.html}} and then flux calibrated using \texttt{IRAF} \citep{iraf1,iraf2}.
Further optical spectroscopy was obtained using the FLOYDS spectrograph on Faulkes Telescope North (FTN) through the Global Supernova Project.
Spectra were reduced with standard methods using a custom \texttt{IRAF}-based pipeline \citep{2014FLOYDS}.
Adding to our high-cadence spectroscopic coverage, we observed SN~2023ixf with the Alhambra Faint Object Spectrograph (ALFOSC) on the Nordic Optical Telescope (NOT; Proposal 67-112, PI: Bonanos).
Observations were reduced with standard reduction techniques using \texttt{IRAF}. 
We add publicly available ALFOSC observations from the NUTS2 collaboration\footnote{The Nordic optical telescope Unbiases Transients Surveys 2; \url{https://nuts.sn.ie}} \citep{2023NOTData}.
In addition, we observed SN~2023ixf with the Multi-Object Double Spectrographs \citep{MODS} on the Large Binocular Telescope (LBT). Data were bias and flat-field corrected using the {\sc modsCCDred} package \citep{pogge_rwpoggemodsccdred_2019}, then extracted and flux calibrated with \texttt{IRAF}.
Spectroscopic observations were taken with the Boller and Chivens Spectrograph (B\&C) on University of Arizona's Bok 2.3m telescope located at Kitt Peak Observatory. These observations were reduced using standard \texttt{IRAF} reduction techniques. 
An optical spectrum was also taken with the Low Resolution Spectrograph 2 \citep[LRS2,][]{LRS2} on the Hobby-Eberly Telescope \citep[HET,][]{HET-1, HET-2} at McDonald Observatory on 2023-06-02. The  data from the red and blue arms (LRS2-R and LRS2-B) were combined into a single spectrum covering the spectral region from 3600 to 10500 \AA. The Integral Field Unit (IFU) spectra were reduced with the \texttt{Panacea} pipeline\footnote{https://github.com/grzeimann/Panacea}.
We collected one 30-minute exposure with the Astrophysical Research Consortium Echelle Spectrograph (ARCES) with resolution $R\approx 31,000$ on the ARC 3.5 m Telescope at Apache Point Observatory (APO). We reduced the spectra with \texttt{IRAF} and \texttt{aesop} \citep{Morris2018}. 
Individual orders normalized using a Spline function.
Further optical spectra were obtained with the 1.22-m Galileo telescope+B\&C at the Asiago Astrophysical Observatory, Italy, which were reduced using an IRAF-based pipeline and the Intermediate Dispersion Spectrograph (IDS) at the Isaac Newton Telescope (INT) which were reduced with the custom python package IDSRED \citep{tomas_e_muller_bravo_2023_7851772}.
.
To this dataset we add publicly available reduced spectroscopic observations from the Liverpool Telescope (LT) archive \citep[SPRAT;][]{2023LTData, 2004LT, 2014SPRAT} and Transient Name Server\footnote{https://www.wis-tns.org/object/2023ixf} (HFOSC, SPRAT). 

A complete list of spectroscopic observations is given in \autoref{sec:SpecApp} and shown in \autoref{fig:montage} and \autoref{fig:montage2}.
Spectra will be made available on Wiserep\footnote{https://www.wiserep.org} \citep{2012wiserep}.

%
%
%

\begin{figure*}
\figurenum{1.1}
\includegraphics[width=\textwidth]{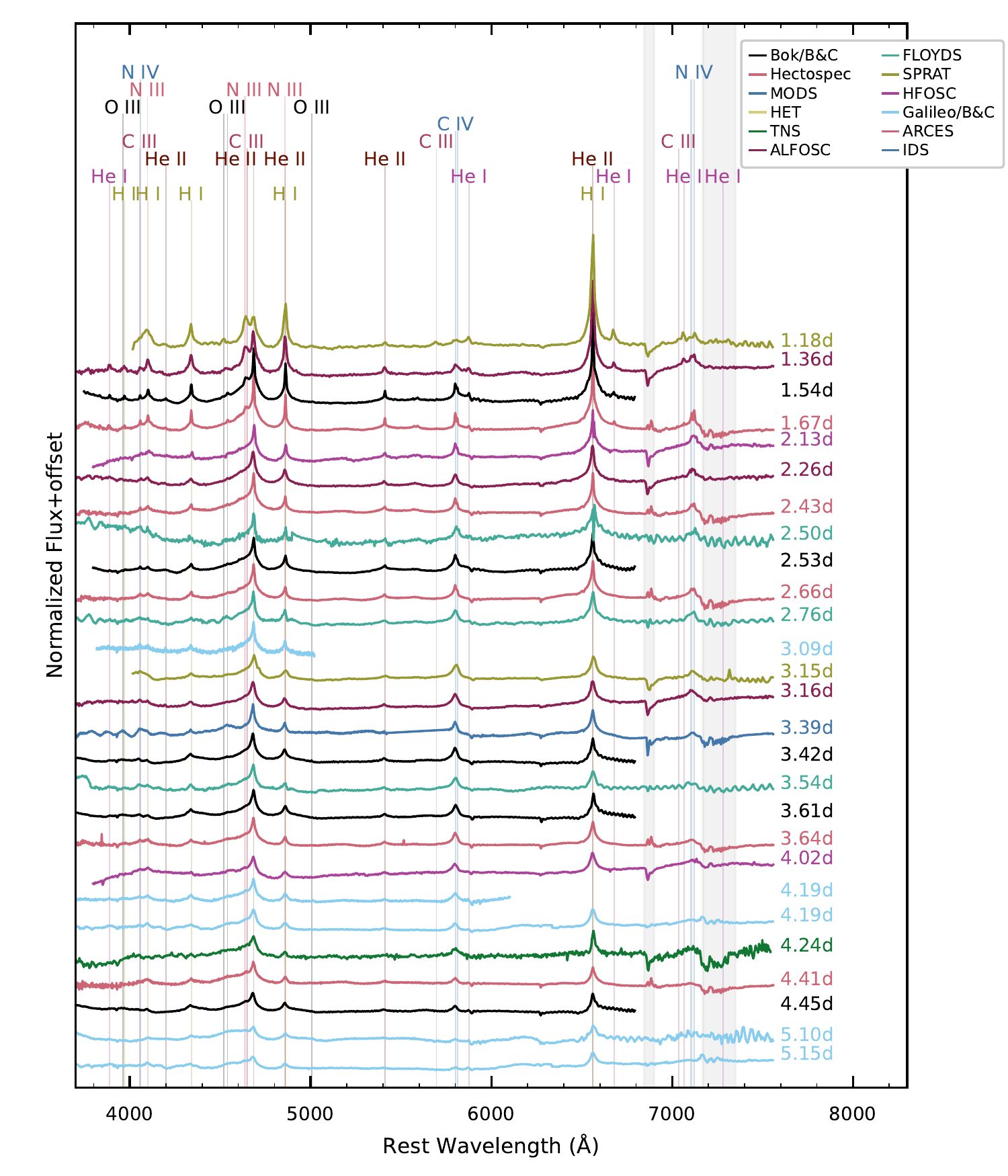}
\caption{The evolution of the optical spectra of SN~2023ixf over the first $\sim5.2$ days, with the earliest epoch at the top of the figure.
Spectra are color coded by the instrument used to observe them, normalized by a black body fit, and corrected for redshift.
Emission lines are identified at their rest wavelengths with vertical lines and labeled at the top of the figure, while the most prominent telluric features are marked with the shaded gray region.
Throughout this sequence, the spectra evolve from showing strong, narrow emission lines from high-ionization species to a intermediate width features as the CSM is accelerated by the shock.
The spectra evolve rapidly between 1.18--1.67d, with the \ion{He}{1} and \NC{} lines disappearing within the first 0.5d from the first spectrum while \HeII{}, \ion{N}{4}, and \ion{C}{4} gain strength. }
    \label{fig:montage}
\end{figure*}
\begin{figure*}
\figurenum{1.2}
\includegraphics[width=\textwidth]{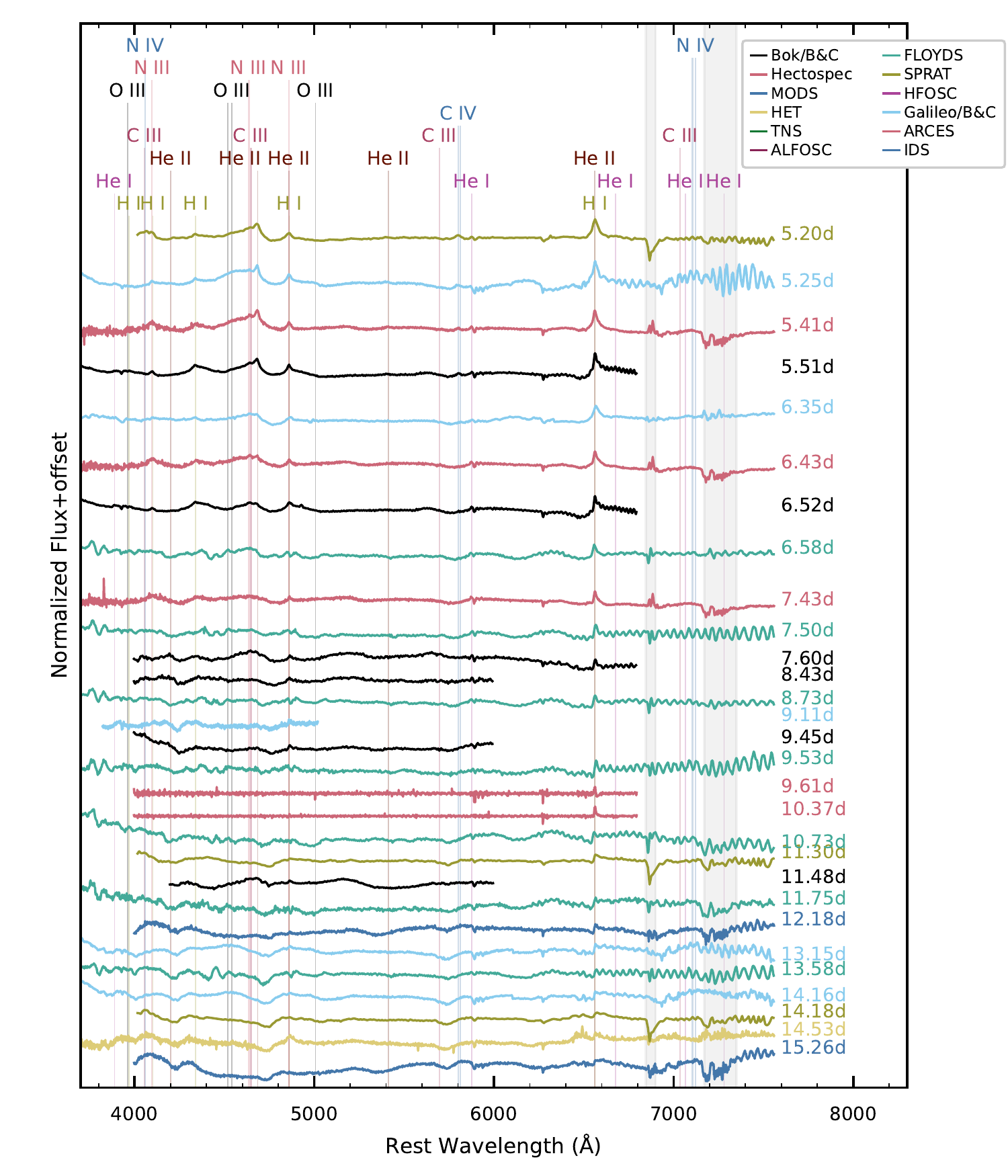}
\caption{Continuation of \autoref{fig:montage} showing the evolution of the optical spectra of SN~2023ixf from day 5.2 to 14.5 with the earliest epoch at the top of the figure.
Over the first 7 days, the spectra evolve from showing strong, narrow emission lines from high-ionization species to a nearly featureless spectrum with an intermediate with P~Cygni profile in \Halpha{}.
In the subsequent 7 days, broad P~Cygni profiles develop in the higher order Balmer features.
}
    \label{fig:montage2}
\end{figure*}
\setcounter{figure}{1}

\section{Fundamental Supernova Parameters}\label{sec:params}
The nearby host galaxy of SN~2023ixf, M101, also hosted the Type Ia SN~2011fe.
We adopt the distance modulus to M101, derived using the Leavitt Law, from \citet{2022riess}: $\mu = 29.178 \pm 0.041$ mag (D=6.85 Mpc).
For Milky Way extinction, we use the SN location and the dust maps of \citet{schlafly_measuring_2011}\footnote{https://irsa.ipac.caltech.edu/applications/DUST/index.html} to find $E(B-V) = 0.0077 \pm 0.0002$ mag. 
Measuring the equivalent width of the \ion{Na}{1}~D lines in high-resolution observations and using the relationship of \citet{Poznanski2012}, \citet{2023Smith} find an average host $E(B-V) = 0.031$ mag with $\pm$30\% uncertainty from the relation between Na~{\sc i} D and extinction; a value which is consistent with other \ion{Na}{1} D extinction measurements made with high resolution data \citep{lundquist_host_2023}.

M101 is a popular target for both amateur and professional observers, and images of the galaxy taken by amateur astronomers \citep{mao_first_det_2023} prior to discovery provided the last deep non-detection (2023-05-18 15:50:24; MJD 60082.66) and first detection (2023-05-18 20:29; MJD 60082.85), with $\lesssim$5 hours separating them.
Following \citet{2023hosseinzadeh}, we define the explosion epoch as half way between the last non-detection and first detection: 2023-05-18 18:00:00 (MJD 60082.75 $\pm$ 0.10), where the adopted uncertainty is the span between the explosion epoch and last non-detection (or first detection).

\section{Spectroscopic Evolution}\label{sec:SpecEvolve}
Tracking the rapid evolution of SN~2023ixf, we observed SN~2023ixf at least four times per night for the first 5 days and at least nightly thereafter. 
While high-cadence spectra have been obtained for a select few SNe over the first few days of evolution \citep{Yaron17, Terreran22}, SN~2023ixf is the first to have intra-night observations for the first week.
Over the first two weeks, the spectra evolve from strong, narrow emission lines with broad wings, to a nearly featureless spectrum and finally develop Balmer P~Cygni profiles with shallow absorption, more typical of the early evolution of Type II SNe. 

To identify spectral features, we use the second spectrum, taken 1.36 days after explosion, as it has higher signal-to-noise and resolution than the classification spectrum. 
From this spectrum, we identify \ion{H}{1}, \ion{He}{1}, \ion{He}{2}, \ion{C}{3}, \ion{C}{4}, \ion{O}{3}, \ion{N}{3}, and \ion{N}{4} lines. 
A full list of species is given in \autoref{sec:IonApp}. 

\begin{figure*}
    \centering
    \includegraphics{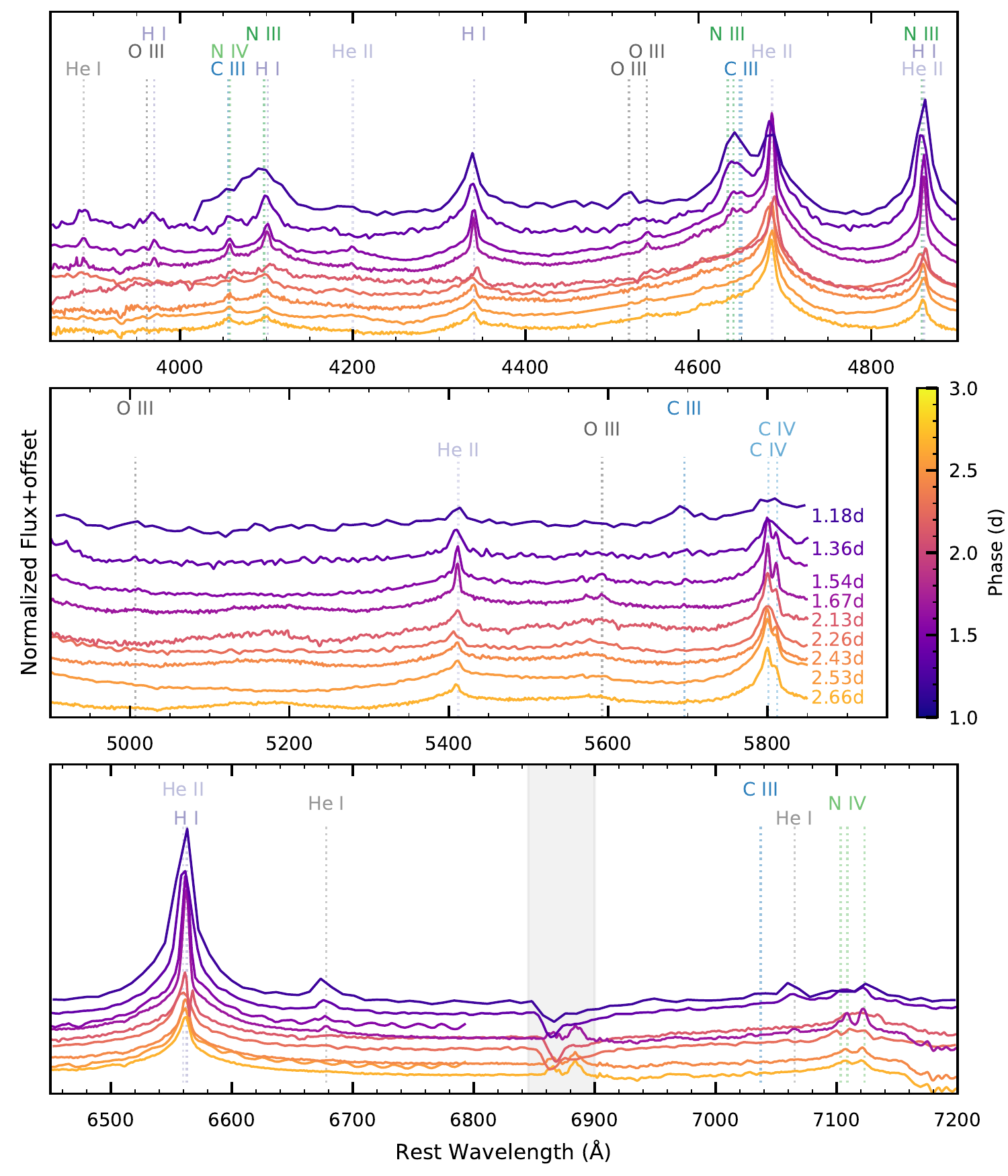}
    \caption{The evolution of SN~2023ixf from day 1.18 through day 3, colored by phase. 
    Ions are labeled in each panel and the phase is given in the middle panel. 
    Emission lines from the low-ionization levels, \ion{He}{1}, \ion{N}{3}, \ion{C}{3}, \ion{O}{3}, and \ion{N}{4} ($\lambda$~5074~\AA{}), disappear over the first 0.5 days of evolution while emission from high-ionization levels, \ion{N}{4} ($\lambda\lambda 7103, 7109, \lambda 7122$), \HeII{}, and \ion{C}{4} ($\lambda\lambda$ 5801.3, 5811.98 \AA{}) increase in strength.
    Spectra are fit with a blackbody function and normalized, corrected for redshift, and offset for readability.
    In the bottom panel, the B-band telluric feature is marked with a gray shaded region.}
    \label{fig:earlyMontage}
\end{figure*}

We see a rapid evolution in the first 0.5 days of our spectra (1.18--1.67 days), which is shown in detail in \autoref{fig:earlyMontage}.
First we turn to the complex of lines around 4700 \AA{} in the top panel of the figure. 
Over this epoch, the blend of \NC{} fades into the broad blueshifted wing of \HeII{} while the latter line increases in strength. 
Similarly, the \ion{He}{1} lines visible in the first spectrum rapidly fade until they are no longer visible 0.5d later.
Other lines such as \ion{O}{3}, \ion{C}{3}, and \ion{N}{4} ($\lambda$ 5074 \AA{}) that are marginally detected in the first few spectra also disappear on this time scale.
At the same time, \ion{N}{4} ($\lambda\lambda 7103, 7109, \lambda 7122$) along with \HeII{}, \ion{C}{4} ($\lambda\lambda$ 5801.3, 5811.98) increase in strength.

The most persistent features through day 5 are \ion{H}{1} ($\lambda$ 4101.73) and possibly \ion{N}{3} ($\lambda$ 4097.33; although it is not possible to determine if two distinct lines exist at this resolution), \HeII{}, and \ion{C}{4} ($\lambda\lambda5801.3, 5811.98$). 
Interestingly, \NC{} become visible again as a broad shelf in the \HeII{} blue wing around day 4, as the \HeII{} line fades. 

Over this time, the strength of all these narrow features decreases to a nearly featureless spectrum 7 days after explosion. 
The one line still clearly present in the spectrum after day 6.5 is \Halpha{}, which appears to develop an asymmetric emission profile, reminiscent of a P~Cygni profile with shallow absorption.
The intermediate width of this feature indicates the presence of CSM that has been accelerated by the shock. 
Details of its evolution are presented by \citet{2023Smith}.
Around day 12, clear broad P~Cygni profiles from the SN ejecta are present in the high-order Balmer features, although the profile at \Halpha{} is significantly more complex with no clear absorption component.

\subsection{Comparison to other Flash SNe}\label{sec:otherflash}
\begin{figure*}
    \centering
    \includegraphics{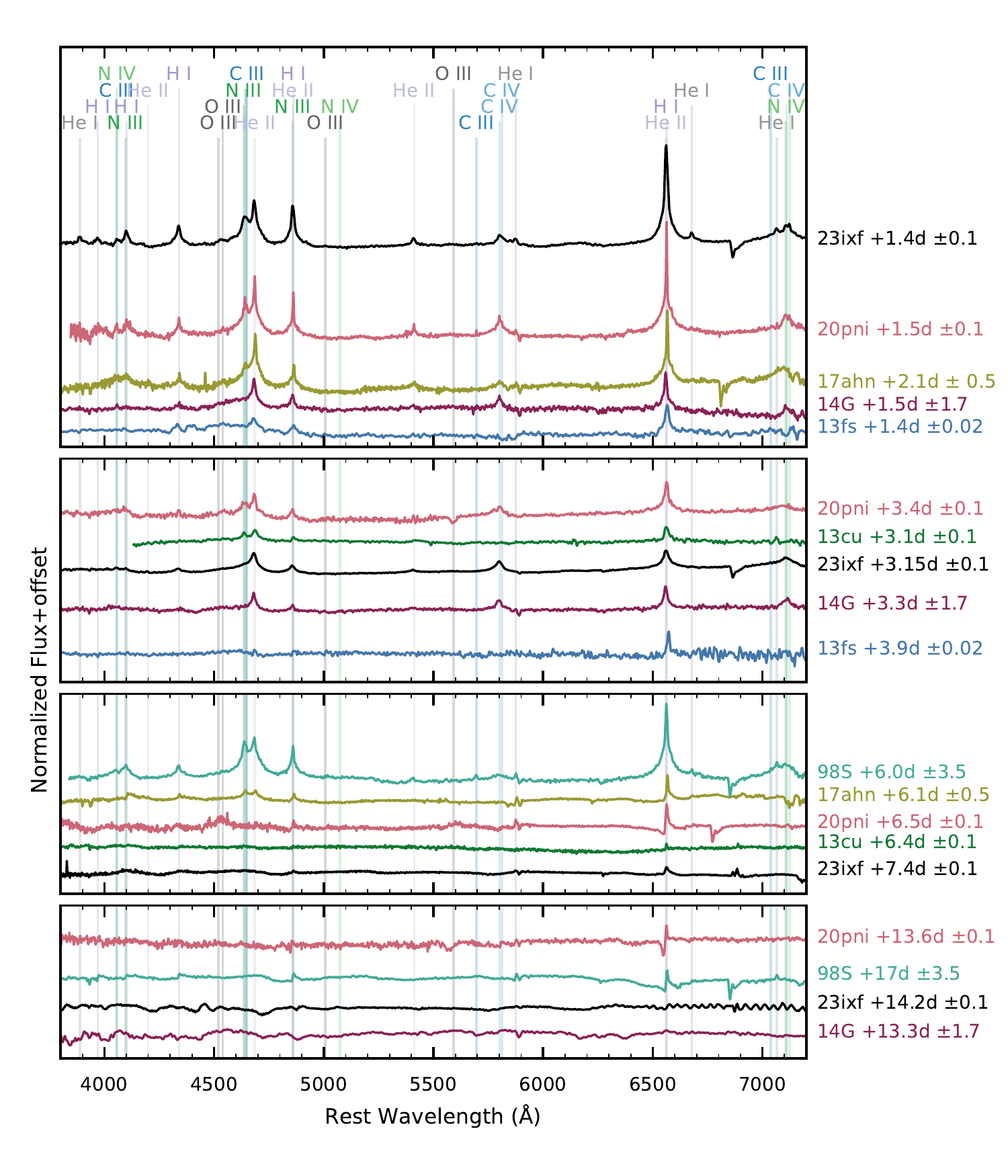}
    \caption{A comparison of SN~2023ixf (black) to SN~1998S \citep[teal;][]{2000leonard}, SN~2013fs \citep[blue;][]{Yaron17}, SN~2013cu \citep[green;][]{GalYam14}, SN~2014G \citep[purple;][]{2016terreran}, SN~2017ahn \citep[mustard;][]{Tartaglia21}, and SN~2020pni \citep[pink;][]{Terreran22} $\sim$1, 3, 7, and 14 days after explosion. 
    These SNe all have flash features that evolve over the first two weeks. 
    The SN name and epoch are marked to the right of the plot.
    All spectra have been redshift and extinction corrected and then normalized by a blackbody fit to the continuum.
    Emission lines identified in the first spectrum are marked with vertical lines which are labeled at the top of the figure. 
    In the first epoch, SN~2023ixf closely resembles SNe~2020pni and 2017ahn, although at later epochs it more closely follows SN~2014G.
    SN~2013fs is considerably different in its evolution throughout the first two weeks. 
    At later epochs, SN~2020pni and SN~1998S develop narrow P~Cygni profiles that are not present in the spectra of other SNe.
    }
    \label{fig:SNComp}
\end{figure*}

A number of SNe have been observed to have narrow emission lines early in their evolution, often disappearing within the first week. 
In \autoref{fig:SNComp} we compare the spectra of SN~2023ixf at day 1, 3, 7, and 14 to those of SN~1998S \citep{2000leonard}, SN~2013fs \citep{Yaron17}, SN~2013cu \citep{GalYam14}, SN~2014G \citep{2016terreran}, SN~2017ahn \citep{Tartaglia21}, and SN~2020pni \citep{Terreran22} at comparable phases.

Given the rapid evolution within the first week of explosion, the uncertainty in the explosion epoch of each comparison supernova should be considered. 
SN~1998S was discovered on 1998-03-02 16:19:12 with the last non-detection 7 days earlier on 1998-02-23 16:47:24. 
With the discovery image taken 18 days before V band maximum, \citet{2000leonard}, conclude that it was likely discovered within a few days of shock breakout. 
We take the explosion epoch to be half way between the last non-detection and first detection (1998-02-27 04:33:18) with an uncertainty of the distance to the last non-detection/first detection: 3.5d.
SN~2013cu was discovered on 2013-05-03 04:19:12. 
\citet{GalYam14} derive an explosion epoch of 2013-05-02 22:19:12 with an uncertainty of 0.11d which we adopt in this paper.
Similarly, \citet{Yaron17} find an explosion epoch of 2013-10-06 02:52:48, a few hours prior to the first detection on 2013-10-06 06:05:45 with an uncertainty of 0.5 hr which we use in this analysis.
\citet{2016terreran} set the explosion epoch of SN~2014G to 2014-01-12 14:24:00, half way between the last non-detection and first detection, with an uncertainty of 1.7d.
SN~2017ahn was discovered on 2017-02-08 06:57:00, 1 day after the last non-detection on 2017-02-07 05:31:12. 
\citet{Tartaglia21} find the explosion epoch to be half way between the last non-detection and first detection: 2017-02-07 18:14:24 with an uncertainty of 0.5d.
For SN~2020pni, \citet{Terreran22} determine an explosion epoch of 2020-07-15 19:12:00, 0.4d prior to the last non-detection, with an uncertainty of 0.1d.

These SNe encompass a range of subtypes of Type II supernovae which are defined by their light curve shape or spectral characteristics. 
Type IIP/L SNe are defined by hydrogen in their spectra throughout their evolution and a plateau from peak brightness to $\sim$100 days after explosion which is either flat or linearly declining, with a continuum of possible slopes.
One explanation for the different slopes is that steeper slopes represent progenitor stars with smaller hydrogen envelopes due to mass loss \citep{1971grassberg, 1989young, 1993blinnikov, 2014anderson, 2016moriya, 2021hiramatsu}.
In the cases where only a very small hydrogen envelope remains, the supernova becomes a Type IIb SN, defined by the transition of their spectra from hydrogen dominated to showing strong helium lines \citep{1997filippenko}. 
All of the comparison SNe are Type IIP/L SNe with the exception of SN~2013cu, which is a Type IIb \citep{2000leonard,GalYam14,2016terreran,Yaron17,Tartaglia21,Terreran22}. 
Their light curves show similarly shallow linear declines in V-band, with SN~2013cu being the steepest \citep{2023yamanaka, 2023teja, 2023jacobson, 2023hosseinzadeh, 2023hiramatsu, 2023arXiv230705612G}.

Among the SNe in this sample, SN~2013fs is the most distinct.
It showed the highest ionization species (e.g. \ion{O}{5}; $\lambda5541$ and \ion{O}{6}; $\lambda\lambda3811, 3834$; \citealt{Yaron17}) in the earliest spectra that are not present in SN~2023ixf or other SNe in this sample. 
However, it is possible that these features were present at earlier times in SN~2023ixf and, by the first spectrum at day 1.18, had disappeared. 
While there are emission lines from some of the same ions in the day 1.5 spectra of SN~2013fs and SN~2023ixf, many are not present in the SN~2013fs spectrum, especially between 3800--4500~\AA{} and 5000--6500 \AA{}, and \ion{N}{5} ($\lambda$ 4604) is absent from the SN~2023ixf spectrum.
Additionally, the emission in all lines is weaker in SN~2013fs and fades more rapidly. 

SN~2023ixf more closely resembles the other comparison objects in \autoref{fig:SNComp}, although it does not always match their evolution.
The first spectrum closely resembles that of SNe~2020pni and 2017ahn. 
However, some features, e.g. \NC{}, are persistent at day 3 in SNe~2020pni and 2017ahn, but are not visible in SN~2023ixf.
This likely indicates that the CSM of SN~2023ixf is cooling more slowly than that of SN~2020pni or SN~2017ahn, perhaps due to a lower density.
At day 3, we also have spectra from SN~2013cu and SN~2014G.
At this phase, SN~2023ixf is best matched to SN~2014G, a trend that continues throughout the two week evolution, while SN~1998S and SN~2013cu more closely resemble SN~2020pni and SN~2017ahn.
At day 7, most features have faded and \Halpha{} has been replaced by an intermediate width P~Cygni profile in SN~2023ixf, in contrast to the narrow P~Cygni profiles in SNe~2020pni and 1998S.
Finally, at day 14, SN~2020pni and SN~1998S still show narrow P~Cygni profiles, while SN~2023ixf and SN~2014G are starting to develop broad P~Cygni profiles.
The disappearance of these features can be interpreted as the ejecta enveloping the CSM, which would indicate that the radial extent of the CSM of SN~2023ixf is smaller than that of SN~2020pni and SN~1998S or the ejecta velocity higher.

\section{Comparison to Models}\label{sec:models}
Variations in mass-loss rate, SN luminosity, surface abundance, and CSM density profile can all affect the characteristics of the narrow emission line spectrum and its evolution.
In the following section, we compare our spectroscopic dataset to two sets of publicly available model grids \citep{Boian20, 2017dessart}, which vary different parameters, using the best-fit model to characterize the progenitor, CSM, and SN properties.
We fit a blackbody to both the model and observed data and normalize by this before comparing them. 
This removes any temperature continuum effects, instead only examining flux relative to the continuum.

\subsection{Boian \& Groh models}
\citet{2019boian} propose that the narrow features produced by the CSM around Type II SNe can be used to constrain the abundances of this material and therefore the mass of the progenitor system. 
They predict different SN line diagnostics for low-mass RSG progenitors (8--15 \msun{}), massive RSG, yellow hypergiant, blue supergiant progenitors (15--30~\msun{}), and stripped stars like LBV and N-rich Wolf-Rayet progenitors (15--30 \msun{}) for observations from one to a few days after explosion.  
These progenitors are then modeled for high-, medium-, and low-luminosity SNe using the radiation transport code CMFGEN \citep{hillier_treatment_1998, hillier_time-dependent_2012, dessart_type_2013, hillier_photometric_2019}.
The primary difference between the low- and high-mass progenitors is in the abundances of the surface material.
Low-mass progenitors should experience weak or no CNO processing. 
High mass progenitors, on the other hand, are expected to have significant CNO processed materials.
Finally, stripped stars should be He-rich.
\autoref{tab:boian} gives the predicted spectroscopic signatures of low and high-mass progenitors and different luminosity SNe.

We search for these features in our 1.36d spectrum.
We see no \ion{O}{6} ($\lambda\lambda 3811, 3834$) which rules out low-mass, high-luminosity SNe as this is the only diagnostic predicted for this combined mass and luminosity.
We also do not see \ion{C}{3} ($\lambda 5697$), which is one of three diagnostics of the low-mass, medium-luminosity system.
It is unclear how to interpret a scenario in which only a subset of the diagnostic lines are present.
Additionally, we identify \ion{N}{4} ($\lambda4058$, $\lambda\lambda 7109,7122$), \ion{N}{3} ($\lambda\lambda4634, 4640$), \ion{C}{3} ($\lambda\lambda4647, 4650$), and \ion{C}{4} ($\lambda\lambda$ 5801, 5811), which are features of the remaining progenitor mass - luminosity combinations. 
Thus, although we can eliminate a low mass, high luminosity progenitor, our conclusions about the remaining combinations of mass and luminosity are inconclusive as the diagnostic lines for multiple scenarios are observed and for some of them only a subset of the features are present.
We note that this diagnostic depends on uncertain physical parameters in single star evolutionary models such as mixing efficiency, mass-loss rates, and convective overshoot which complicate the connection between surface abundance, as measured from CSM, and initial mass.
Additionally, the temperature of the CSM has a dramatic effect on the ions present, independent of surface abundances, which is a function of both the individual supernova parameters as well as the phase at which the model and observations are compared.
This further complicates the connection between surface abundances and progenitor masses.

As the broad trends noted by \citet{2019boian} were inconclusive, we visually compare their full set of model spectra for all mass-loss rates, luminosities, and progenitors to our 1.36d spectrum. 
The clear presence of both \ion{N}{3} ($\lambda$4634) and \ion{He}{2} ($\lambda$4685.5) with \ion{He}{2} stronger than \ion{N}{3} greatly limits the number of possible models to $L=1.5\times10^{9}$~\lsun{} (which they define as medium luminosity) and $\dot{M}=3\times10^{-3}$ \msun{} yr$^{-1}$ (these lines are not present in any other models). 
The wind velocity of these models is $v_{w}=150~\mathrm{km\,s^{-1}}$ which is consistent with the observed CSM velocity \citep{2023Smith}.

While we have identified a mass-loss rate and luminosity which is consistent with our observed spectrum based on the presence of individual features, \citet{Boian20} scale the model luminosity to match the observed luminosity, which leads to a scaling in the mass-loss rate as well.
We repeat this analysis to derive a luminosity and mass-loss rate for SN~2023ixf. 
First, \citet{Boian20} identify the temperature via the ionization level present.
Similarly, we selected the best models based on continuum normalized spectra, looking only at the relative ionization levels.
In their model, the temperature and luminosity are related via the Stefan-Boltzmann law.
Thus after identifying the temperature, they scale to the observed luminosity, scaling the radius to maintain the derived temperature.
They also derive a scale factor for the mass-loss rate, assuming that $\dot{M}\propto L_{SN}^{3/4}$.
We duplicate this analysis for SN~2023ixf for the three surface abundances (which do not produce significantly different results), concluding that the mass-loss rate of the progenitor of SN~2023ixf was $\dot{M}\approx4.5\times10^{-3}$ \mlunit{} and scaled luminosity for each abundance is $L\approx2.6\times10^{9}$~\lsun{}.

\begin{figure*}[t]
    \centering
    \includegraphics[width=\textwidth]{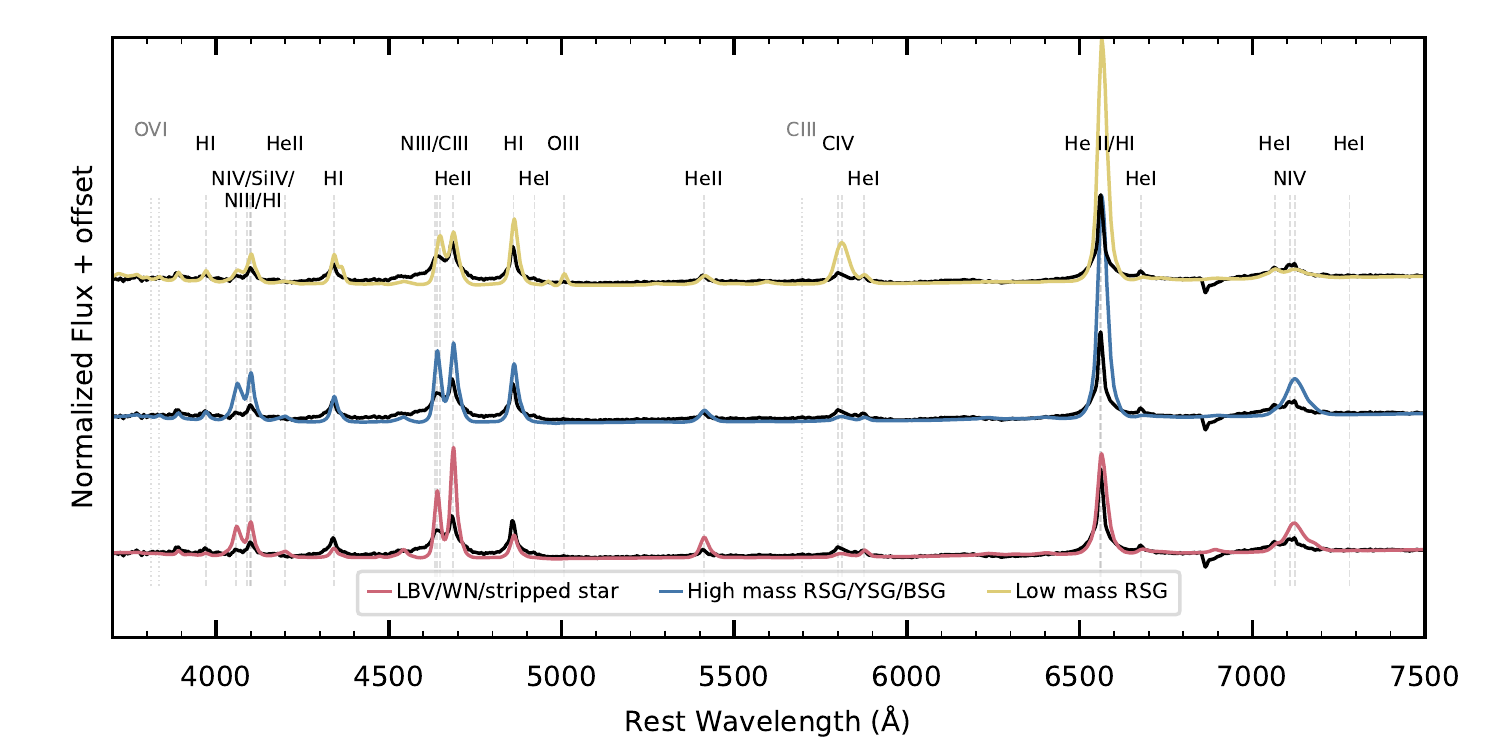}
    \caption{The day 1.36 spectrum (black) compared with three different surface abundance models from \citet{2019boian} with $L=1.5\times10^{9}$ \lsun{} and $\dot{M}=3\times10^{-3}$ \msun{}$~\mathrm{yr^{-1}}$. 
    The solar abundance represents a low-mass RSG (yellow), CNO-processed abundance represents a high-mass RSG/BSG/YSG (blue), and He-rich abundance represents a LBV/WN/stripped star (pink).
    Vertical dashed lines represent the ions that are identified in the observed spectra that are labeled at the top of the plot. 
    Vertical dotted lines in light gray show ions that are not detected but are part of the progenitor diagnostics detailed in \citet{2019boian}.
    We convolve the model spectra with a Gaussian kernel to mimic the resolution of the observed spectra. 
    While some individual features in each model are matched by the observed spectrum, there are differences between the model and observed spectrum that indicate that the conditions of the model are not the same as the observations.}
    \label{fig:boian}
\end{figure*}

Having constrained the luminosity and mass-loss rate, we examine the different surface abundances: the solar abundance spectra, corresponding to the low mass RSG scenario; the CNO-processed surface abundance, corresponding to the high-mass RSG, BSG, YSG scenario; and the He-rich abundance, corresponding to the LBV, WN, stripped star scenario.
These are compared to an observed spectrum in \autoref{fig:boian}.
We find that no one scenario matches the line strengths of the observation. 
While many of the lines are well matched in the solar abundance model, the \ion{C}{4} ($\lambda\lambda5801, 5811$) and \Halpha{} are greatly over-estimated by the model.
The \ion{C}{4} ($\lambda\lambda5801, 5811$) is better represented in the CNO abundance model (a result of the suppression of C and O), however, the N in our observed spectrum is significantly weaker than the model, countering the expectation that these stars would be nitrogen enriched.
Finally, in the He-rich model, the \ion{H}{1} line is well-modeled and the \HeII{} line is stronger in the model than in the observation. 
Like the CNO abundance model, the N lines are too strong in the model.  
It is possible that, rather than indicating surface abundance, these discrepancies arise from a mismatch between the physical and model CSM density, temperature, and/or the hardness of the radiation field.

\begin{deluxetable*}{ccc}
\tablecaption{Predicted spectroscopic signatures from \citet{2019boian}. \label{tab:boian}}
\tablehead{\colhead{Luminosity} & \colhead{Low-mass Progenitor (8--15 \msun{})} & \colhead{High-mass Progenitor (15--30 \msun{})} }
\startdata
&&\\
Low luminosity  & \ion{C}{3} ($\lambda\lambda4647, 4650$)&  \ion{C}{4} ($\lambda\lambda5801, 5811$) \\
($\sim1.9\times10^{9}$ \lsun{})& & Lack of \ion{C}{3} ($\lambda5697$)\\
&&\\
\hline
&&\\
Medium luminosity  & \ion{C}{3} ($\lambda$5697)& \ion{N}{4} ($\lambda4058$, $\lambda\lambda7109, 7122$)\\
($3.9\times10^{8}-3.1\times10^{9}$ \lsun{})&  \ion{C}{4} ($\lambda\lambda5801, 5811$) & \\
& \ion{N}{3} ($\lambda\lambda4634, 4640$) & \ion{N}{3} ($\lambda\lambda4634, 4640$)\\
&&\\
\hline
&&\\
High luminosity ($>6.3\times10^{9}$ \lsun{}) & \ion{O}{6} ($\lambda\lambda3811, 3834$) &  Lack of \ion{O}{6} ($\lambda\lambda3811, 3834$)\\
&&\\
\enddata
\end{deluxetable*}

\subsection{Dessart \& Hillier models}
In another study, \citet{2017dessart} model the spectroscopic signature of CSM interaction with a variety of RSG mass-loss rates ($\dot{M}=10^{-6}-10^{-2}$ \msun{}~$\mathrm{yr^{-1}}$) and atmospheric density scale heights ($H_{\rho}=0.01, 0.1, 0.3  R_{*}$). 
The base of each RSG model is a 15~\msun{} star onto which they add an atmosphere with a given density scale height, which transitions to wind mass loss when the density of the atmosphere equals the density of the wind. 
The wind is then extended to $R_{\mathrm{out, CSM}}=5\times10^{14}$ cm for all but one model which is extended to $R_{\mathrm{out, CSM}}=2\times10^{14}$ cm. 
At this point all models transition to an $\dot{M}=10^{-6}$ \msun{}\,$\mathrm{yr^{-1}}$ wind.
The parameters of each model are summarized in \autoref{tab:dessart}.
Each of these models is evolved from shock breakout to over 10 days and snapshots of the spectra and light curves are reported.

Given the uncertainties in the explosion epoch for the observations and challenges of explosion in the models (e.g. core-collapse vs shock breakout, varying shock breakout time scales in dense CSM), we compare our observed spectra with the full suite of models for all mass-loss rates and epochs. 
In the weak-wind models (r1w1, r1w1h, r2w1), the only spectrum in the time series with narrow emission features is the first spectrum at shock break out. 
As our lines are clearly present throughout the first 5--7 days of evolution, we do not examine these models further.
With this cut alone, we constrain the mass-loss rate to be $\dot{M}>10^{-3}$\mlunit{}.
The remaining models (r1w4, r1w6, r1w5r, and r1w5h) show multiple epochs of narrow emission features.
These features give way to narrow P~Cygni profiles in the \HeII{} and \Halpha{} lines which are eventually replaced by broader features originating from the bulk motion of the ejecta.

\begin{figure*}
    \centering
    \includegraphics[width=\textwidth]{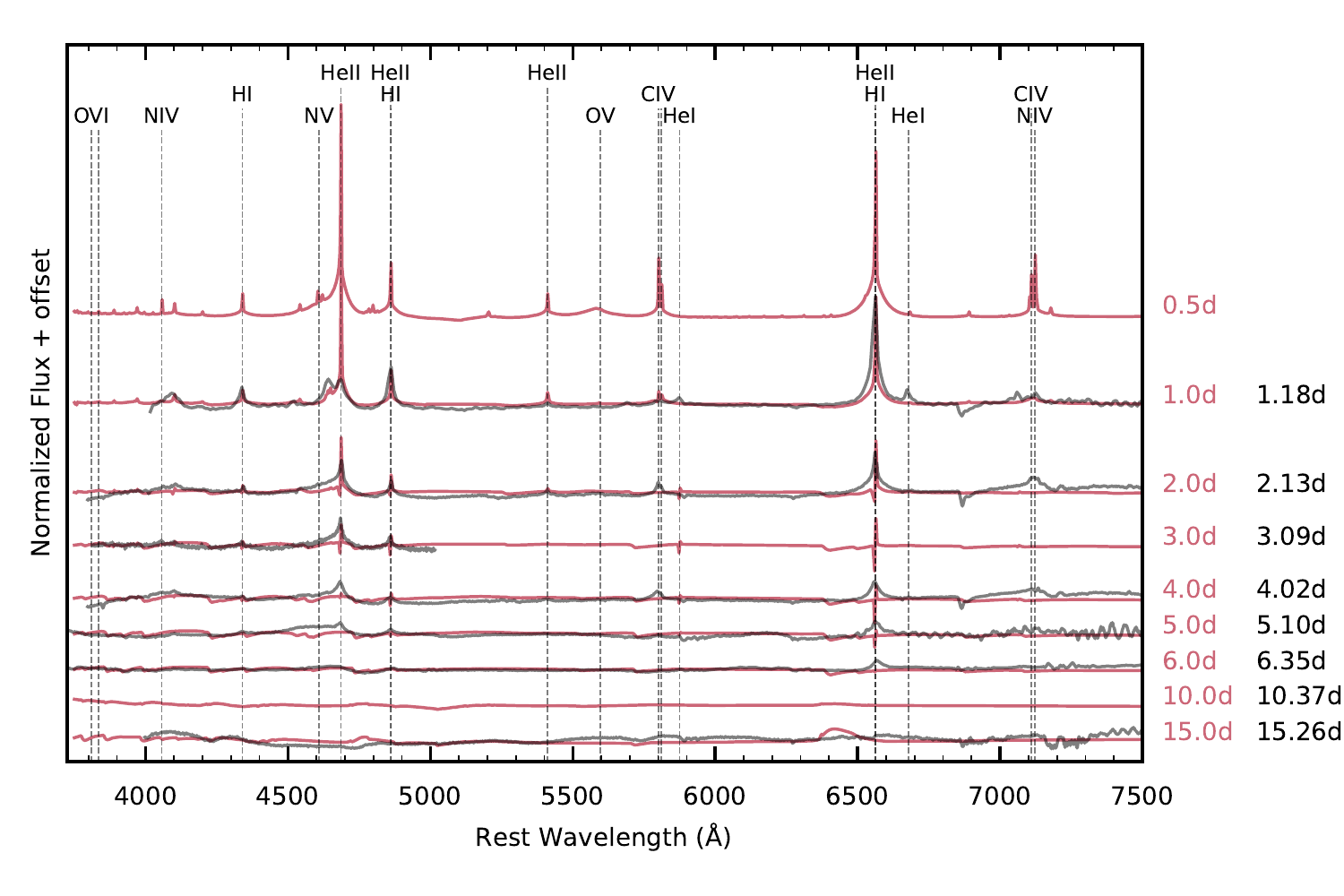}
    \caption{A comparison of the spectral evolution of SN~2023ixf (black) to the r1w4 model of \citet{2017dessart} (pink). Phase are shown to the right of the figure in pink for the model and black from the observed spectra. While the first observation at 1.2 day matches fairly well, the model spectra evolve much more rapidly, with P~Cygni profiles developing at day~2 and all emission disappearing by day 4, while the observations show emission through day 6-7.
    The model spectrum at 0.5d shows a number of lines that have faded by 1d (e.g. \ion{N}{5}). }
    \label{fig:dessart}
\end{figure*}

Again, none of the models reproduces the observed spectra.
Broadly, the emission features in the model spectra fade much more rapidly than in the observed spectra, if they are present at all.
This implies that the CSM in SN~2023ixf extends beyond that of the models or has a higher density (e.g. the r1w6 model is the only one with narrow emission lines that persist past 2 days).
Particularly challenging is the blended \NC{} complex and the \ion{He}{1} lines, which are clearly visible in the observed spectrum at day 1.36 and quickly fades below detection by day 2.0. 
These lines are not present in most of the model spectra at any epoch. 
Additionally, the spectra of SN~2023ixf never show \ion{N}{5} ($\lambda4610$), which is present in the early spectra of all of the strong wind models, although in some models this has faded by the phase of our first spectrum (e.g. r1w4 in \autoref{fig:dessart}).
Instead, we see \ion{N}{3} ($\lambda4636$), which has a similar flux to \ion{He}{2} ($\lambda4686$) in the classification spectrum and then fades over the subsequent 0.5d. 
Although the r1w4 and r1w6 models show this feature just prior to the development of the narrow P~Cygni features, it is always significantly weaker than the \ion{He}{2} emission.
Given the rapid evolution of this feature, this could be a function of the model sampling.
We note that we do not see evidence of the rise of \ion{N}{5} described by \citet{2023jacobson} in our spectra.
Rather, we find the asymmetric blue wings of \HeII{} to be more consistent with \NC{}.

We find the best agreement with the r1w4 and r1w6 models which, while not able to reproduce the observed line ratios, show \ion{N}{3}, \ion{He}{1} ($\lambda7065$), \ion{C}{4} ($\lambda\lambda5801, 5811$), and \ion{N}{4} ($\lambda4057$, $\lambda7122$) features. 
Although present, the \ion{C}{4} and \ion{N}{4} are significantly stronger in the first model spectra than in the observed spectra.
Additionally, the strength of the Balmer emission lines is better matched in these models as is the lack of the \ion{O}{5} ($\lambda5597$).
\autoref{fig:dessart} shows the spectral evolution of SN~2023ixf compared to the r1w4 model.
While the first spectrum matches well, the narrow lines disappear from the model by day 2 and narrow P~Cygni profiles emerge.
In the final spectrum, the model shows significantly more \Halpha{} emission than the observation.
Interestingly, the mass-loss rate of this model is $10^{-3}$~\msun{}\,$\mathrm{yr^{-1}}$, consistent with the conclusions from the comparison to the models of \citet{2019boian}.

The first r1w6 model spectrum on day 1.3 clearly shows \ion{N}{5}.
Additionally, r1w6 model time series does not show the \ion{N}{3} shoulder on \ion{He}{2} until 1-2 days after it has disappeared from the observed spectra, however, the features are otherwise reasonably matched. 
The disappearance of the narrow emission lines and evolution of \Halpha{} are better matched in the r1w6 model.
This is consistent with the findings of \citet{2023jacobson}, who find a best fit model in their custom grid is the r1w6 model with a larger radius.
Given the challenges of reproducing the observed spectra with published models, we conservatively conclude that the mass-loss rate for the progenitor of SN~2023ixf was between 10$^{-3}$ and 10$^{-2}$ \mlunit{}, based on the presence of persistent lines after the initial shock breakout spectrum.
The spectra themselves are sensitive to density, which is parameterized as $\dot{M}=\rho v_{w}$, thus a different mass-loss rate is inferred if the assumed velocity is different.
The wind velocity in these models is $v_{w}=50~\mathrm{km\,s^{-1}}$, which is about a factor of 3 smaller than the measured velocity. 
However, given the order of magnitude range in mass-loss rate inferred from these models, we do not further modify the mass-loss rates.

\begin{deluxetable*}{ccccc}
\tablecaption{CSM Parameters for the Models of \citet{2017dessart} \label{tab:dessart}}
\tablehead{\colhead{Model Name} & \colhead{Mass-loss Rate (\mlunit{})}  & \colhead{Radius (\rsun{})}  & \colhead{Scale Height ($R_{\star}$)}& Transition Radius$^{\star}$ (cm)}
\startdata
r1w1 & $10^{-6}$ & 501 & 0.01 & $5\times10^{14}$ \\
r1w2 & $10^{-5}$ & 501 & 0.01 & $5\times10^{14}$ \\
r1w3 & $10^{-4}$ & 501 & 0.01 & $5\times10^{14}$ \\
r1w4 & $10^{-3}$ & 501 & 0.01 & $5\times10^{14}$ \\
r1w5 &  $5\times10^{-3}$ & 501 & 0.01 & $5\times10^{14}$ \\
r1w6 & $10^{-2}$ & 501 & 0.01 & $5\times10^{14}$ \\
r1w1h & power law$^{\dagger}$ & 501 & 0.3 & $5\times10^{14}$ \\
r1w5r & $10^{-5}$ & 501 & 0.01 & $2\times10^{14}$ \\
r2w1 &  $10^{-6}$ & 1107 & 0.01 & $5\times10^{14}$ \\
\enddata
\tablecomments{~$^{\dagger}$power-law exponent with an exponent of 12}
\tablecomments{~$^{\star}$Radius at which the density transitions to $\dot{M}=10^{-6}$ \mlunit{}}
\end{deluxetable*}

\section{Model-independent CSM Constraints}\label{sec:CSM}
The properties and evolution of the narrow emission lines allow us to compute order of magnitude estimates of the CSM characteristics.
A lower limit on the outer CSM radius can be calculated from the brightest \Halpha{} flux \citep[$F_\mathrm{H\alpha}$;][]{Yaron17, 2013ofek}.
Briefly, the \Halpha{} luminosity ($L_\mathrm{H\alpha}$) is produced by the recombination of ionized H. 
Assuming a spherically symmetric CSM that is composed of hydrogen, all of which is ionized, we can use the \Halpha{} luminosity to calculate the total CSM mass.  
The total mass can also be calculated by integrating a constant velocity wind density profile over the radial extent of the CSM. 
Equating these two relations allows us to determine the radius of the CSM:
\begin{equation}\label{eqn:Rmin}
r\gtrsim\frac{\kappa L_\mathrm{H\alpha}}{A}
\end{equation}
where $\kappa=0.34~\mathrm{cm^{2}\,g^{-1}}$ is the electron scattering opacity of the CSM.
We use the distance $D$ to convert $F_\mathrm{H\alpha}$ to $L_\mathrm{H\alpha}$ via $L_\mathrm{H\alpha}=4\pi D^2F_\mathrm{H\alpha}$.
A is defined as
\begin{equation}
A =\frac{4\pi h\nu_{H} \alpha_{H}^\text{eff}}{\mu_{p} m_{p}^{2}}
\end{equation}
where 
$\nu_{H}=4.56\times10^{14}~\mathrm{Hz}$ is the frequency of \Halpha{},
$\alpha_{H}^\text{eff}=8.7\times10^{14}~\mathrm{cm^{3}\,s^{-1}}$ is the H recombination coefficient for case B recombination at $T_\text{eff}=10000~\mathrm{K}$, 
$\mu_{p}=0.5$ is the mean molecular weight, $m_{p}$ is the proton mass, and $h$ is Planck's constant.
We treat \autoref{eqn:Rmin} as a lower limit on the outer CSM radius, as either the composition or ionization assumption may not be true, which would lead to a larger CSM radius than calculated here.
Additionally, we assume that the CSM above the emitting region is transparent to the \Halpha{} photons.
If this is not true, it would also lead to a larger CSM radius.

To measure the flux of \Halpha{}, we fit a blackbody to the continuum and subtract it from the flux of our 1.36 day spectrum, which has the maximum \Halpha{} flux of our spectral series.
To the continuum-subtracted flux, we simultaneously fit broad and narrow Lorentzian emission profiles.
We integrate this fit from 6300--6800 \AA{} to find $F_\mathrm{H\alpha}=3.18\times10^{-13}~\mathrm{erg\,cm^{-2}\,s^{-1}}$.
Using a distance of 6.85 Mpc, we find $L_\mathrm{H\alpha}=1.78\times10^{39}~\mathrm{erg\,s^{-1}}$ and $R_\text{CSM, out}\gtrsim8.7\times10^{13}~\mathrm{cm}$.
Assuming a spherical wind with a constant velocity, mass-loss rate, and homogeneous density structure, the density can be calculated from the radius:
\begin{equation}
    \rho = \frac{1}{\kappa \, r}
\end{equation}
From this, we calculate a density of $3.4\times10^{-14}~\mathrm{g\,cm^{-3}}$.
Assuming a typical RSG wind of $v_{w}\approx10~\mathrm{km\,s^{-1}}$, this mass loss event would have begun $\gtrsim3~\mathrm{yr}$ before explosion.  
However, using the mass-loss rate derived from high-resolution spectroscopy in \citet{2023Smith} of $v_{w}\approx150~\mathrm{km\,s^{-1}}$ we find a much smaller start time of $\sim 2$ months prior to explosion.

The narrow features in the CSM are only present when there is unshocked, photoionized CSM in front of the SN shock and ejecta.
Therefore, we expect these features to disappear when the material producing them is swept up by the ejecta, and we can use this information to calculate the radius of the CSM. 
In SN~2023ixf, the unshocked, narrow features disappear from \Halpha{} 3--4 days after explosion. 
\citet{2023Smith} find this corresponds to a radius of $R=(3-5)\times10^{14}$ cm.
However, at this phase, there is still material in front of the photosphere which produces intermediate width lines and eventually an intermediate width P~Cygni profile in \Halpha{}.
These intermediate width lines disappear around day $\sim$6--7. 
On day 13, clear broad absorption is visible in H$\beta$ and from this we approximate an average ejecta velocity of $\sim9000~\mathrm{km\,s^{-1}}$.
Putting this together with the time that the lines disappear, we calculate a CSM radius of $R_\mathrm{CSM, out}\sim5.4\times10^{14}~\mathrm{cm}$.
Interestingly, this is exactly where the models of \citet{2017dessart} transition to $\mathrm{\dot{M}=10^{-6}}$ \mlunit{}.
Again, assuming a constant RSG wind of $v_{w}\approx10~\mathrm{km\,s^{-1}}$ (or $v_{w}\approx150~\mathrm{km\,s^{-1}}$), this implies that the event began $\sim 17\,\mathrm{yr}$ ($1\,\mathrm{yr}$ for $v_{w}\approx150~\mathrm{km\,s^{-1}}$) before explosion.
Assuming a constant wind density profile, the density is related to radius by:
\begin{equation}
    \rho=\frac{\dot{M}}{4\pi v_{w} R_{\mathrm{CSM}}^{2}}
\end{equation}
Using a representative mass-loss rate of $5\times10^{-3}$ \mlunit{}, we find a density of $\rho=8.5\times10^{-14}~\mathrm{g\,cm^{-3}}$ ($\rho=5.6\times10^{-14}~\mathrm{g\,cm^{-3}}$ for $v_{w}\approx150~\mathrm{km\,s^{-1}}$).
This is consistent with the lower limit calculated from the \Halpha{} luminosity.

We compare the mass-loss rates derived in \autoref{sec:models} to a population of Type II SNe with flash ionization features to identify how unusual (or normal) these values are. 
We use the sample of \citet{Boian20}, who use CMFGEN to model a sample of 17 Type II SNe that show narrow emission features in their early spectra, indicating CSM interaction.
In \autoref{sec:models}, we used a spectrum taken 1.36 days after explosion to determine the mass-loss rate and SN luminosity for SN~2023ixf. 
Given our extensive time sampling, we investigate whether the epoch of the spectrum used to do this would affect which model was selected. 
We tested this using spectra taken on days 2.26, 3.16, and 5.15.
Given the uncertainty in the epoch of the models, we consider all model epochs for each observed spectrum.
We find that, in all cases, regardless of the epoch used, we would select the models and thus mass-loss rates from \autoref{sec:models}.
Additionally, we find that with the day 5.15 spectrum, we would include the r1w5h model, which has a mass-loss rate between the two models we selected (r1w4 and r1w6).
This demonstrates that the differences between the models are robust to observed epoch as long as narrow features are present.

\begin{figure*}
    \centering
    \includegraphics[width=\textwidth]{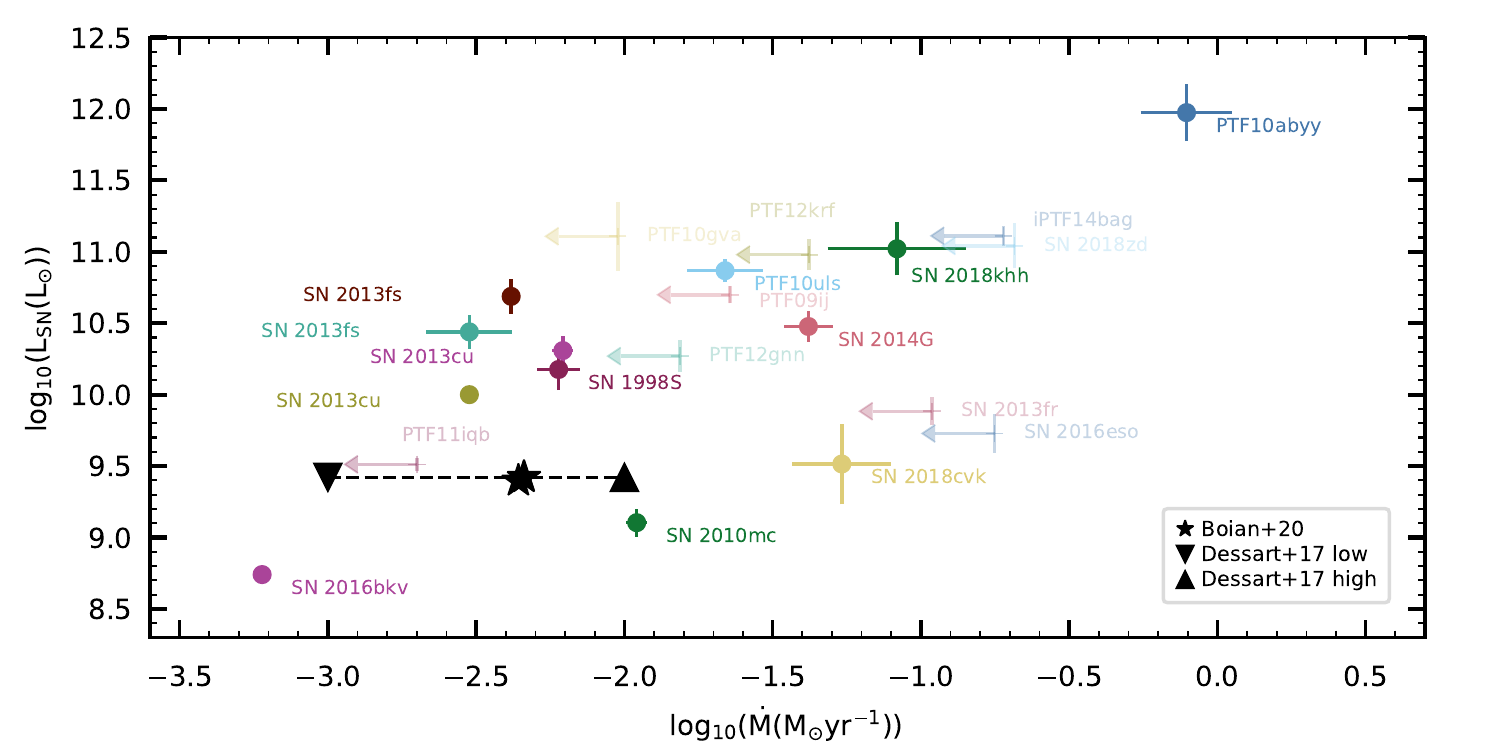}
    \caption{The SN luminosity and mass-loss rate derived by \citet{Boian20} for a sample of 17 SNe compared to the best model of SN~2023ixf (black).
    Upper limits on mass loss are shown with semi-transparent markers and arrows, while determined values are solid. 
    We determine a scaled mass-loss rate of $\dot{M}\approx4.5\times10^{-3}$ \mlunit{} and a scaled luminosity of $L\approx2.6\times10^{9}$\lsun{} from the models of \citet{2019boian} (with slight variation due to different surface abundances) which we plot as a black stars. 
    In practice, these are located at virtually the same location on this plot.
    We also include the mass-loss rates of the r1w4 and r1w6 models of \citet{2017dessart} ($\dot{M}=10^{-3}$ \mlunit{} and $\dot{M}=10^{-2}$ \mlunit{}, respectively) which are shown as black triangles.
    \citet{2017dessart} does not explore variations in luminosity and we therefore use the luminosity derived from \citet{2019boian} solar abundance models and do not scale the mass-loss rates. 
    The mass-loss rate of SN~2023ixf is in line with the lower end of the range of mass-loss rates in this sample.}
    \label{fig:boiansample}
\end{figure*}

With this confirmation, we add SN~2023ixf to Figure~8 from \citet{Boian20}, which shows the SN luminosity and mass-loss rate for all SNe in their sample (\autoref{fig:boiansample}).
We find the mass-loss rate of SN~2023ixf to be fairly low when compared to the range of mass-loss rates for other SN with early interaction, although it is in no way an outlier. 
This is consistent with a lower density CSM causing the persistence of the higher-ion levels of \ion{N}{4} and \ion{C}{4}.

While in line with other interacting events, the mass-loss rates determined for the confined CSM in this paper are significantly higher than steady-state RSG winds.
Using the progenitor luminosity $\log(L/$\lsun{}$\mathrm{)=4.94}$ found by \citet{2023jencson} and the \citet{1988dejaeger} relation, the expected mass-loss rate would be $\dot{M}\approx10^{-5.7}$ \mlunit{} and even lower for the progenitor luminosity identified in \citet{2023kilpatrick}. 
Using the relationship of \citet{2020beasor} and the stellar parameters of \citet{2023jencson}, the expected mass-loss rate would be $\dot{M}\approx2\times10^{-6}$ \mlunit{}.
Even using the enhanced mass-loss rates of \citet{2012ekstrom}, the mass-loss rate would only be $\dot{M}\approx10^{-5.2}$ \mlunit{}.
This implies that the mass-loss event which led to this CSM was vastly greater than the nominal RSG mass loss.
On the other hand, the largest mass-loss rate that we find is consistent with the lower range of mass-loss rates required by \citet{2018morozova} (assuming a 10 $\mathrm{km\,s^{-1}}$ wind) to fit the rapid light curve rises seen in Type II SNe. 

We make a final note on the first 0.5d evolution of the \NC{} emission line.
If the CSM were suddenly ionized by a high energy flash from shock breakout photons, then one would expect
that the CSM is almost instantaneously (modulo the light travel time) ionized. 
Thus, as time passes in the days after the initial flash ionization, the spectral evolution should proceed from high ionization to low ionization species as the initially highly ionized CSM recombines over a timescale determined by the density of the CSM.
However, we observe the opposite. 
Spectra of SN~2023ixf show that relatively low-ionization levels are present in the earliest epochs (i.e. narrow \NC{} and \ion{He}{1} emission on day 1-2). 
Over the following day, these fade away as other higher ionization features (such as \HeII{}, N~{\sc iv}, C~{\sc iv}) strengthen. 
This is also discussed by \citet{2023Smith} and \citet{2023jacobson}.
This implies  that we are witnessing a gradual or delayed ionization of the CSM, an effect never observed before. 
This effect seems inconsistent with a sudden flash ionization from shock breakout alone, and more indicative of a slowly varying source of ionization (i.e. radiation from the CSM interaction shock itself or asymmetry in the ejecta).

\section{Summary \& Conclusions}\label{sec:Conclusion}
The comprehensive spectroscopic data set presented in this paper provides sub-day cadence spectroscopic observations for the first week of evolution and at least daily cadence through day 14.
To our knowledge, no other SN with flash ionization features has been studied with this cadence for this length of time.
The unique combination of cadence and duration of these observations will enable the community to trace the CSM density profile, and therefore the mass-loss history of the progenitor, in a way that has never before been possible.

These spectra show narrow high-ionization emission features with Lorentzian wings.
We identify emission from \ion{H}{1}, \ion{He}{1}, \ion{He}{2}, \ion{C}{3}, \ion{C}{4}, \ion{O}{3}, \ion{N}{3}, and \ion{N}{4}. 
\ion{He}{1}, \ion{C}{3}, \ion{N}{3}, and \ion{O}{3} fade over our first 0.5d of monitoring ($\sim$1.5-2d since explosion).
We find the spectrum at 1.36d most closely resembles that of SNe~2017ahn and 2020pni and looks significantly different from SN~2013fs. 
Over time, the evolution closely matches that of SN~2014G, as the high-ionization features fade and eventually so does \ion{He}{2} and \ion{H}{1}. 
The differences in the evolution of these different SNe imply that SN~2023ixf has a lower density CSM than SNe~2017ahn and 2020pni and a smaller radial extent.
By day 13, broad P~Cygni profiles have developed indicating emission from the SN ejecta.

We compare the same 1.36d spectrum with the models of \citet{2019boian}, finding the spectrum most closely resembles the models with $L\approx$2.6$\times10^{9}$ \lsun{} and $\dot{M}\approx$4.5$\times10^{-3}$ \mlunit{}, although we do not find any model that reproduces the line ratios in our observed spectra and therefore cannot use these diagnostics to infer a progenitor mass.

We also relate the full spectral evolution over the first two weeks to the models of \citet{2017dessart}, which examine different mass-loss rates and atmospheric scale heights. 
We find the spectra of SN~2023ixf are best represented by the r1w4 and r1w6 models, corresponding to mass-loss rates of $\mathrm{\dot{M}=10^{-3}}$ \msun{} $\mathrm{yr^{-1}}$ and $\mathrm{\dot{M}=10^{-2}}$ \msun{} $\mathrm{yr^{-1}}$ respectively.
However, we note that in r1w4 model, the narrow emission lines disappear much more rapidly from the model than we observe indicating that a larger radial extent is required in the model.
We find that despite the rapid evolution of the spectrum over the first five days, the spectra are most consistent with the same models, regardless of epoch used to identify them.

Finally, we use the narrow lines to calculate the properties of the CSM. 
Using the maximum \Halpha{} flux and assuming a spherical geometry, we find $R_\text{CSM, out}\gtrsim8.7\times10^{13}~\mathrm{cm}$, implying the CSM ejection began at least 3 years ago if it is expanding at 10 $\mathrm{km\,s^{-1}}$ and 2 months ago if the observed velocity of $150~\mathrm{km\,s^{-1}}$.
At this radius, we find a CSM density of $3.3\times10^{-14}~\mathrm{g\,cm^{-3}}$.
Using the epoch at which time the narrow emission features disappear, we find a consistent radius of $R_\text{CSM}=5.4\times10^{14}~\mathrm{cm}$.
With this radius and RSG wind $v_{w}\approx10~\mathrm{km\,s^{-1}}$ ($v_{w}\approx150~\mathrm{km\,s^{-1}}$), the CSM ejection began 17 years ago (1 year ago) and the density is $\rho=8.5\times10^{-14}~\mathrm{g\,cm^{-3}}$ ($\rho=5.6\times10^{-14}~\mathrm{g\,cm^{-3}}$). 
Comparing SN~2023ixf to a sample of 17 SNe with early CSM interaction, we find the mass-loss rate in line with the lower end of the distribution and the low luminosity. 
We note that this analysis assumes spherically symmetric CSM. 
Asymmetric CSM (such as that proposed by \citet{2023Smith} or clumped CSM \citep[e.g.][]{dessart_impact_2018} would alter these conclusions, although the details of the effect will depend on the exact configuration.

SN~2023ixf is an extraordinary SN, combining proximity with early detection and classification, and tight constraints on explosion.
The immediate announcement of the discovery and classification allowed us to harness our resources and observe the detailed evolution of the CSM interaction over the first two weeks. 
With these observations we identify a significantly higher mass-loss rate than the nominal RSG mass-loss rate of $\mathrm{\dot{M}=10^{-6}}$ \msun{} $\mathrm{yr^{-1}}$. 
This indicates either a superwind or period of eruptive mass loss \citep[but see also][]{2019kochanek}.
While both the models of \citet{2019boian} and \citet{2017dessart} are unable to match the temporal evolution and the relative flux ratios, the majority of the emission lines present are reproduced and we find the prospects of a custom model to match the observations encouraging.
The spectroscopic data set presented in this paper can help guide future modeling efforts and be used to benchmark the evolution flash features in any SN.

\section*{Acknowledgments}
We thank our anonymous referee for providing comments to improve the paper.
This publication was made possible through the support of an LSSTC Catalyst Fellowship to K.A.B., funded through Grant 62192 from the John Templeton Foundation to LSST Corporation. The opinions expressed in this publication are those of the authors and do not necessarily reflect the views of LSSTC or the John Templeton Foundation.
Time domain research by the University of Arizona team and D.J.S.\ is supported by NSF grants AST-1821987, 1813466, 1908972, \& 2108032, and by the Heising-Simons Foundation under grant \#20201864. 
The research by Y.D., S.V., N.M., and E.H.\ is supported by NSF grants AST-2008108.
J.E.A.\ is supported by the international Gemini Observatory, a program of NSF's NOIRLab, which is managed by the Association of Universities for Research in Astronomy (AURA) under a cooperative agreement with the National Science Foundation, on behalf of the Gemini partnership of Argentina, Brazil, Canada, Chile, the Republic of Korea, and the United States of America.
The research of J.C.W.\ and J.V.\ is supported by NSF AST-1813825. 
J.V.\ is also supported by OTKA grant K-142534 of the National Research, Development and Innovation Office, Hungary.
L.S.\ and M.W.C.\ acknowledge support from the National Science Foundation with grant numbers PHY-2010970 and OAC-2117997. 
A.Z.B\ acknowledges support from the European Research Council (ERC) under the European
Union’s Horizon 2020 research and innovation programme (Grant agreement No. 772086).
AR acknowledges support from ANID BECAS/DOCTORADO NACIONAL 21202412.
We thank the MMT director, G.~Williams, for granting Director's Discretionary Time for the Hectospec spectral sequence.  This paper made use of the modsCCDRed data reduction code developed in part with funds provided by NSF Grants AST-9987045 and AST-1108693. 
A.P. and P.O. acknowledge support of the PRIN-INAF 2022 project ``Shedding light on the nature of gap transients: from the observations to the models''.
The SNICE research group 
acknowledges financial support from the Spanish Ministerio de Ciencia e Innovaci\'on (MCIN), the Agencia Estatal de Investigaci\'on (AEI) 10.13039/501100011033, the European Social Fund (ESF) ``Investing in your future'', the European Union Next Generation EU/PRTR funds, the Horizon 2020 Research and Innovation Programme of the European Union, and by the Secretary of Universities and Research (Government of Catalonia), under the PID2020-115253GA-I00 HOSTFLOWS project, the 2019 Ram\'on y Cajal program RYC2019-027683-I, the 2021 Juan de la Cierva program FJC2021-047124-I, the Marie Sk\l{}odowska-Curie and the Beatriu de Pin\'os 2021 BP 00168 programme, and from Centro Superior de Investigaciones Cient\'ificas (CSIC) under the PIE project 20215AT016, and the program Unidad de Excelencia Mar\'ia de Maeztu CEX2020-001058-M.

We thank David Bohlender, Dmitry Monin, and James Di Francesco for obtaining DAO spectra and the whole LBT team, especially Alexander Becker and Jennifer Power.
Based on observations obtained with the Hobby-Eberly Telescope (HET), which is a joint project of the University of Texas at Austin, the Pennsylvania State University, Ludwig-Maximillians-Universitaet Muenchen, and Georg-August-Universitaet Goettingen. The HET is named in honor of its principal benefactors, William P.\ Hobby and Robert E.\ Eberly. The Low Resolution Spectrograph 2 (LRS2) was developed and funded by the University of Texas at Austin McDonald Observatory and Department of Astronomy, and by Pennsylvania State University. We thank the Leibniz-Institut fur Astrophysik Potsdam (AIP) and the Institut fur Astrophysik Goettingen (IAG) for their contributions to the construction of the integral field units.
The Liverpool Telescope is operated on the island of La Palma by Liverpool John Moores University in the Spanish Observatorio del Roque de los Muchachos of the Instituto de Astrofisica de Canarias with financial support from the UK Science and Technology Facilities Council.
The data presented here were obtained in part with ALFOSC, which is provided by the Instituto de Astrofisica de Andalucia (IAA) under a joint agreement with the University of Copenhagen and NOT.
Based on observations made with the Nordic Optical Telescope, owned in collaboration by the University of Turku and Aarhus University, and operated jointly by Aarhus University, the University of Turku and the University of Oslo, representing Denmark, Finland and Norway, the University of Iceland and Stockholm University at the Observatorio del Roque de los Muchachos, La Palma, Spain, of the Instituto de Astrofisica de Canarias.
This project has received funding from the European Union’s Horizon 2020 research and innovation programme under grant agreement No 101004719 (ORP: OPTICON RadioNet Pilot).
Observations reported here were obtained at the MMT Observatory, a joint facility of the University of Arizona and the Smithsonian Institution.
Based on observations made with the 1.22-m Galileo Galilei Telescope of the Padova University in the Asiago site.
This paper made use of the modsCCDRed data reduction code developed in part with funds provided by NSF Grants 
AST-9987045 and AST-1108693. 

\facilities{ADS, ARC (ARCES), Asiago:Galileo (B\&C), Bok (B\&C), HCT (HFOSC), HET (LRS2), INT (IDS), LBT (MODS), LCOGT (FLOYDS), Liverpool:2m (SPRAT), MMT (Hectospec),  NED, NOT (ALFOSC), TNS}

\software{ aesop \citep{Morris2018}, astropy \citepalias{astropy_collaboration_astropy_2013, astropy_collaboration_astropy_2018, astropy_collaboration_astropy_2022}, CMFGEN \citep{hillier_treatment_1998, hillier_time-dependent_2012, hillier_photometric_2019, dessart_type_2013}, FLOYDS \citep{valenti_first_2014}, HSRED, IRAF \citep{iraf1, iraf2}, Light Curve Fitting \citep{hosseinzadeh_light_2023}, MatPLOTLIB \citep{hunter_matplotlib_2007},  MODS pipeline \citep{pogge_rwpoggemodsccdred_2019}, NumPy \citep{harris_array_2020}, Scipy \citep{virtanen_scipy_2020}. }

\appendix
\section{Spectroscopic Observations}\label{sec:SpecApp}
\autoref{tab:spec} lists the date, telescope, instrument, and resolving power for each spectroscopic observation used in this paper.
\startlongtable
\begin{deluxetable*}{cccccc}
\tablecaption{Log of Spectroscopic Observations \label{tab:spec}}
\tablehead{\colhead{Phase (d)} & \colhead{Time} & \colhead{MJD} & \colhead{Telescope} & \colhead{Instrument}& \colhead{R ($\lambda$/$\Delta \lambda$)}}
\startdata
1.18 & 2023-05-19 22:23:45 & 60083.93 & LT & SPRAT & 350 \\
1.36 & 2023-05-20 02:39:56 & 60084.11 & NOT & ALFOSC & 360 \\
1.54 & 2023-05-20 07:03:24 & 60084.29 & Bok & B\&C & 700 \\
1.67 & 2023-05-20 10:00:43 & 60084.42 & MMT & Hectospec & 1325 \\
2.13 & 2023-05-20 21:04:19 & 60084.88 & HCT & HFOSC & 350 \\
2.26 & 2023-05-21 00:14:57 & 60085.01 & NOT & ALFOSC & 360 \\
2.43 & 2023-05-21 04:19:39 & 60085.18 & MMT & Hectospec & 1325 \\
2.50 & 2023-05-21 06:07:11 & 60085.25 & FTN & FLOYDS & 500 \\
2.53 & 2023-05-21 06:40:49 & 60085.28 & Bok & B\&C & 700 \\
2.66 & 2023-05-21 09:50:44 & 60085.41 & MMT & Hectospec & 1325 \\
2.76 & 2023-05-21 12:07:58 & 60085.51 & FTN & FLOYDS & 500 \\
3.09 & 2023-05-21 20:03:50 & 60085.84 & Galileo & B\&C & 2762 \\
3.15 & 2023-05-21 21:41:41 & 60085.90 & LT & SPRAT & 350 \\
3.16 & 2023-05-21 21:48:37 & 60085.91 & NOT & ALFOSC & 360 \\
3.39 & 2023-05-22 03:21:44 & 60086.14 & LBT & MODS & 2075 \\
3.42 & 2023-05-22 03:57:46 & 60086.17 & Bok & B\&C & 700 \\
3.54 & 2023-05-22 07:00:07 & 60086.29 & FTN & FLOYDS & 500 \\
3.61 & 2023-05-22 08:41:45 & 60086.36 & Bok & B\&C & 700 \\
3.64 & 2023-05-22 09:15:15 & 60086.39 & MMT & Hectospec & 1325 \\
4.02 & 2023-05-22 18:31:41 & 60086.77 & HCT & HFOSC & 350 \\
4.19 & 2023-05-22 22:33:56 & 60086.94 & NOT & ALFOSC & 300 \\
4.19 & 2023-05-22 22:32:09 & 60086.94 & Galileo & B\&C & 1195 \\
4.24 & 2023-05-22 23:50:00 & 60086.99 & Other$^{\dagger}$ & Other$^{\dagger}$ & -$^{\dagger}$ \\
4.41 & 2023-05-23 03:43:23 & 60087.16 & MMT & Hectospec & 1325 \\
4.45 & 2023-05-23 04:47:09 & 60087.20 & Bok & B\&C & 700 \\
5.10 & 2023-05-23 20:16:48 & 60087.85 & Galileo & B\&C & 983 \\
5.15 & 2023-05-23 21:33:06 & 60087.90 & NOT & ALFOSC & 300 \\
5.20 & 2023-05-23 22:48:32 & 60087.95 & LT & SPRAT & 350 \\
5.25 & 2023-05-24 00:01:26 & 60088.00 & Galileo & B\&C & 983 \\
5.41 & 2023-05-24 03:47:15 & 60088.16 & MMT & Hectospec & 1325 \\
5.51 & 2023-05-24 06:07:28 & 60088.26 & Bok & B\&C & 700 \\
6.35 & 2023-05-25 02:18:49 & 60089.10 & NOT & ALFOSC & 300 \\
6.43 & 2023-05-25 04:13:21 & 60089.18 & MMT & Hectospec & 1325 \\
6.52 & 2023-05-25 06:35:59 & 60089.27 & Bok & B\&C & 700 \\
6.58 & 2023-05-25 08:02:16 & 60089.33 & FTN & FLOYDS & 500 \\
7.43 & 2023-05-26 04:15:36 & 60090.18 & MMT & Hectospec & 1325 \\
7.50 & 2023-05-26 05:55:24 & 60090.25 & FTN & FLOYDS & 500 \\
7.60 & 2023-05-26 08:17:43 & 60090.35 & Bok & B\&C & 700 \\
8.43 & 2023-05-27 04:21:08 & 60091.18 & Bok & B\&C & 700 \\
8.73 & 2023-05-27 11:25:38 & 60091.48 & FTN & FLOYDS & 500 \\
9.11 & 2023-05-27 20:35:31 & 60091.86 & Galileo & B\&C & 2455 \\
9.45 & 2023-05-28 04:52:00 & 60092.20 & Bok & B\&C & 700 \\
9.53 & 2023-05-28 06:40:09 & 60092.28 & FTN & FLOYDS & 500 \\
9.61 & 2023-05-28 08:35:03 & 60092.36 & APO & ARCES & 30000 \\
10.36 & 2023-05-29 02:47:20 & 60093.12 & APO & ARCES & 30000 \\
10.73 & 2023-05-29 11:26:52 & 60093.48 & FTN & FLOYDS & 500 \\
11.30 & 2023-05-30 01:11:14 & 60094.05 & LT & SPRAT & 350 \\
11.48 & 2023-05-30 05:24:06 & 60094.23 & Bok & B\&C & 700 \\
11.75 & 2023-05-30 12:00:24 & 60094.50 & FTN & FLOYDS & 500 \\
12.10 & 2023-05-30 20:18:14 & 60094.85 & Galileo & B\&C & 2007 \\
12.18 & 2023-05-30 22:18:38 & 60094.92 & INT & IDS & 1092 \\
12.55 & 2023-05-31 07:06:58 & 60095.30 & FTN & FLOYDS & 500 \\
13.15 & 2023-05-31 21:40:19 & 60095.90 & Galileo & B\&C & 1019 \\
13.58 & 2023-06-01 07:53:58 & 60096.33 & FTN & FLOYDS & 500 \\
14.16 & 2023-06-01 21:51:50 & 60096.91 & Galileo & B\&C & 1037 \\
14.18 & 2023-06-01 22:19:25 & 60096.93 & LT & SPRAT & 350 \\
14.53 & 2023-06-02 06:44:48 & 60097.28 & HET & LRS2 & 600\\
15.26 & 2023-06-03 00:19:09 & 60098.01 & INT & IDS & 1092
\enddata
\tablecomments{~$^{\dagger}$ This spectrum on TNS does not have any information regarding telescope, instrument, or resolving power.}
\end{deluxetable*}

\section{Line Identification}\label{sec:IonApp}
\autoref{tab:ion} gives the ion and wavelength of the lines identified and shown throughout this paper.
\begin{deluxetable}{cD}
\tablecaption{List of ions observed in the day 1.36 spectrum. \label{tab:ion}}
\tablehead{\colhead{Ion} & \colhead{Wavelength (\AA{})}}
\startdata
He I & 3888.64 \\
O III & 3961.59 \\
H I & 3970.0 \\
C III & 4056.0 \\
N IV & 4057.76 \\
N III & 4097.33 \\
H I & 4101.73 \\
He II & 4200.0 \\
H I & 4340.47 \\
O III & 4519.62 \\
O III & 4540.4 \\
N III & 4634.0 \\
N III & 4640.64 \\
C III & 4647.5 \\
C III & 4650.0 \\
He II & 4685.5 \\
N III & 4858.82 \\
He II & 4860.0 \\
H I & 4861.3 \\
O III & 5006.8 \\
He II & 5412.0 \\
O III & 5592.37 \\
C III & 5695.9 \\
C IV & 5801.3 \\
C IV & 5811.98 \\
He I & 5875.6 \\
He II & 6559.8 \\
H I & 6562.7 \\
He I & 6678.15 \\
C III & 7037.25 \\
He I & 7065.19 \\
N IV & 7103.24 \\
N IV & 7109.0 \\
N IV & 7122.98 \\
He I & 7281.34 \\
\enddata
\end{deluxetable}

\bibliography{sn23ixf_ref}{}
\bibliographystyle{aasjournal}

\end{document}